\documentclass[numberedappendix,appendixfloats]{emulateapj}

\usepackage{amssymb}
\usepackage{amsmath}
\usepackage{float}
\usepackage[font=footnotesize,caption=false,labelsep=quad]{subfig}
\usepackage{booktabs}
\usepackage{ulem}
\usepackage[usenames,dvipsnames]{color}
\usepackage{appendix}

\usepackage{xspace}
\usepackage{graphicx}
\usepackage[export]{adjustbox}
\usepackage[colorlinks,breaklinks, citecolor=blue,linkcolor=blue,urlcolor=blue]{hyperref}
\usepackage{apjfonts}
\usepackage{etoolbox}

\newcommand{\spark}{\small{SPARK}\normalsize{}}
\newcommand{\sersic}{S\'{e}rsic }

\def\Msun{\hbox{M$_{\odot}$}}

\def\halpha{H$\alpha$  }

\def\halphans{H$\alpha$} 
\def\hbetans{H$\beta$} 

\def\hbeta{H$\beta$  }

\def\NII{[N\hspace{.03cm}II]}
\def\SII{[S\hspace{.03cm}II]}

\def\OIII{[O\hspace{.03cm}III]}

\def\mdyn{$M_\mathrm{dyn}$}
\def\kms{km s$^{-1}$}
\def\dsfr{$\Delta$SFR}
\def\stot{$\sigma_\mathrm{tot}$}
\def\skpc{$\Sigma_\mathrm{1kpc}$}

\def\kmostd{KMOS$^\mathrm{3D}$~}
\def\kmostdns{KMOS$^\mathrm{3D}$}

\makeatletter
\patchcmd{\NAT@citex}
    {\@citea\NAT@hyper@{\NAT@nmfmt{\NAT@nm}\NAT@date}}
    {\@citea\NAT@nmfmt{\NAT@nm}\NAT@hyper@{\NAT@date}}
    {}
    {}
\patchcmd{\NAT@citex}
    {\@citea\NAT@hyper@{%
         \NAT@nmfmt{\NAT@nm}%
         \hyper@natlinkbreak{\NAT@aysep\NAT@spacechar}{\@citeb\@extra@b@citeb}%
         \NAT@date}}
    {\@citea\NAT@nmfmt{\NAT@nm}%
     \NAT@aysep\NAT@spacechar%
     \NAT@hyper@{\NAT@date}}
    {}
    {}
\patchcmd{\NAT@citex}
    {\@citea\NAT@hyper@{%
         \NAT@nmfmt{\NAT@nm}%
         \hyper@natlinkbreak
         {\NAT@spacechar\NAT@@open\if*#1*\else#1\NAT@spacechar\fi}%
         {\@citeb\@extra@b@citeb}%
         \NAT@date}}
    {\@citea\NAT@nmfmt{\NAT@nm}%
        \NAT@spacechar\NAT@@open\if*#1*\else#1\NAT@spacechar\fi%
        \NAT@hyper@{\NAT@date}}
    {}
    {}
\makeatother

\shorttitle{KMOS$^\mathrm{3D}$ kinematics of compact SFGs}
\shortauthors{Wisnioski et al.}

\begin{document}
\title{The KMOS$^\mathrm{3D}$ Survey:  rotating compact star forming galaxies \\ 
and the decomposition of integrated line widths \footnotemark[$\dagger$]} \footnotetext[$\dagger$]{Based on observations obtained at the Very Large Telescope (VLT) of the European Southern Observatory (ESO), Paranal, Chile (ESO program IDs 092A-0091, 093.A-0079, 094.A-0217,095.A-0047, 096.A-0025, and 097.A-0028, 098.A-0045}
\author{E. Wisnioski\altaffilmark{1,2,3}, J. T. Mendel\altaffilmark{1,2,3}, N. M. F\"orster Schreiber\altaffilmark{1},  R. Genzel\altaffilmark{1,4}, 
D. Wilman\altaffilmark{5,1}, S. Wuyts\altaffilmark{6},~S. Belli\altaffilmark{1},\\
%K. Bandara\altaffilmark{-},
A. Beifiori\altaffilmark{5,1}, 
R. Bender\altaffilmark{5,1}
G. Brammer\altaffilmark{7}, 
%Burkert, A
J. Chan\altaffilmark{8}, 
R. I. Davies\altaffilmark{1},  
R. L. Davies\altaffilmark{1},  
M. Fabricius\altaffilmark{1}, \\ 
M. Fossati\altaffilmark{5,1},
A. Galametz\altaffilmark{1},  
%S. Kulkarni\altaffilmark{-},  
%J. Kurk\altaffilmark{2}, %??
P. Lang\altaffilmark{9}, 
D. Lutz\altaffilmark{1}, 
E. J. Nelson\altaffilmark{1}, 
I. Momcheva\altaffilmark{7},
D. Rosario\altaffilmark{10},\\
R. Saglia\altaffilmark{1}, 
L. J. Tacconi\altaffilmark{1},
K. Tadaki\altaffilmark{11},
H. \"Ubler\altaffilmark{1},
P. G. van Dokkum\altaffilmark{12}
}
%E. Wuyts\altaffilmark{2}}
\altaffiltext{1}{Max-Planck-Institut f\"{u}r extraterrestrische Physik (MPE), Giessenbachstr. 1, D-85748 Garching, Germany}
\altaffiltext{2}{Research School of Astronomy and Astrophysics, Australian National University, Canberra, ACT 2611, Australia}
\altaffiltext{3}{ARC Centre for Excellence in All-Sky Astrophysics in 3D (ASTRO 3D)}
\altaffiltext{4}{Departments of Physics and Astronomy, University of California, Berkeley, CA 94720, USA}
\altaffiltext{5}{Universit\"ats-Sternwarte, Ludwig-Maximilians-Universit\"at, Scheinerstrasse 1, D-81679 M\"unchen, Germany}
\altaffiltext{6}{Department of Physics, University of Bath, Claverton Down, Bath, BA2 7AY, UK}
\altaffiltext{7}{Space Telescope Science Institute, 3700 San Martin Drive, Baltimore, MD 21218, USA}
\altaffiltext{8}{Department of Physics and Astronomy, University of California, Riverside, CA 92521, USA}
\altaffiltext{9}{Max-Planck-Institut f\"{u}r Astronomie, K\"onigstuhl 17, D-69117 Heidelberg, Germany}
\altaffiltext{10}{Center for Extragalactic Astronomy, Department of Physics, Durham University, South Road, Durham, DH1 3LE, UK}
\altaffiltext{11}{National Astronomical Observatory of Japan, 2-21-1 Osawa, Mitaka, Tokyo 181-8588, Japan}
\altaffiltext{12}{Astronomy Department, Yale University, New Haven, CT 06511, USA}
\email{email: emily.wisnioski@anu.edu.au}

\received{3 November 2017}
\accepted{16 February 2018}
\published{12 March 2018}

%%%%%%%%%%%%%%%%%%%%%%%%%%%%%%%%%%%%%%%%%%%%%%%%
\begin{abstract}
%%%%%%%%%%%%%%%%%%%%%%%%%%%%%%%%%%%%%%%%%%%%%%%%
Using integral field spectroscopy we investigate the kinematic properties of 35 massive centrally-dense and compact star-forming galaxies (${\log{\overline{M}_*[\Msun]}}=11.1$, $\log{(\Sigma_\mathrm{1kpc}[\mathrm{\Msun~kpc}^{-2}])}>9.5$, $\log{(M_\ast/r_e^{1.5}[\mathrm{\Msun~kpc}^{-1.5}])}>10.3$) at $z\sim0.7-3.7$ within the \kmostd survey. We spatially resolve 23 compact star-forming galaxies (SFGs) and find that the majority are dominated by rotational motions with velocities ranging from {$95-500$ \kms}.
The range of rotation velocities is reflected in a similar range of integrated \halpha linewidths, $75-400$ \kms, consistent with the kinematic properties of mass-matched extended galaxies from the full \kmostd sample. The fraction of compact SFGs that are classified as `rotation-dominated' or `disk-like' also mirrors the fractions of the full \kmostd sample. We show that integrated line-of-sight gas velocity dispersions from \kmostd are best approximated by a linear combination of their rotation and turbulent velocities with a lesser but still significant contribution from galactic scale winds.  The \halpha exponential disk sizes of compact SFGs are on average $2.5\pm0.2$ kpc, $1-2\times$ the continuum sizes, in agreement with previous work. The compact SFGs have a $1.4\times$ higher AGN incidence than the full \kmostd sample at fixed stellar mass with average AGN fraction of 76\%. Given their high and centrally concentrated stellar masses as well as stellar to dynamical mass ratios close to unity, the compact SFGs are likely to have low molecular gas fractions and to quench on a short time scale unless replenished with inflowing gas. The rotation in these compact systems suggests that their direct descendants are rotating passive galaxies. \\
\end{abstract}


%\clearpage

%%%%%%%%%%%%%%%%%%%%%%%%%%%%%%%%%%%%%%%%%%%%%%%%
\section{Introduction}
\setcounter{footnote}{0}
%%%%%%%%%%%%%%%%%%%%%%%%%%%%%%%%%%%%%%%%%%%%%%%%

%
%{\color{BrickRed}Quenching is an important event in galaxy evolution...}\\
The transformation of a star-forming galaxy (SFG) into a passive galaxy is a key process in galaxy evolution. Yet the details of how SFGs are quenched are mostly unknown, particularly in the early universe. Studies of the high-redshift galaxy population have shown that quiescent galaxies at $z\gtrsim1$ are much more compact than their local counterparts (e.g., \citealt{2008ApJ...677L...5V,2007MNRAS.382..109T,2009ApJ...695..101D,2012ApJ...746..162N}). Their small sizes suggest that they were either created from larger SFGs through highly dissipative processes or formed from already compact SFGs. These different formation histories imply different quenching mechanisms. 

%{\color{BrickRed} Theoretically several scenarios proposed ... but observationally poorly constrained...}\\
%What are the dominant pathways to quenching $-$ particularly in the early universe as galaxies were rapidly building up mass? 
There are many well supported candidates for quenching $-$ e.g. environment, mass $-$ and evidence that different mechanisms act in tandem (e.g. \citealt{1972ApJ...178..623T,1994ApJ...425L..13M,2010ApJ...721..193P,2010ApJ...715..202H,2017ApJ...835..153F}). However because the transition of \textit{individual galaxies} from star forming to passive can be rapid \citep{2004ApJ...615L.101B,2007ApJS..173..342M} it is particularly difficult to observe individual cases. Instead, progress in understanding the cessation of star formation typically comes from studies of statistical samples where comparisons can be drawn between the global properties of star-forming and passive galaxy populations. Global properties, e.g.~color, structural properties, size-mass, star formation rate-mass, often show a bimodality with a small number of galaxies creating a bridge between the star-forming and passive populations. The galaxies in between may represent an interesting class of galaxies currently undergoing a transformation.

%
%{\color{BrickRed} Processes seem to be associated with structural transformation}\\
In recent years, observational studies have shown that the quenching of star-formation is intimately linked to the formation of a bulge (e.g. \citealt{2014MNRAS.441..599B,2014ApJ...788...11L}), lending support to an association between early-type morphology and quiescence. Galaxies in transition are expected to change their morphologies prior to quenching, stimulated for example by interactions with other galaxies or their environments (e.g. \citealt{1972ApJ...178..623T,1994ApJ...425L..13M,2003ApJ...597..893N}).  Major mergers can drive the formation of spheroids either directly or by triggering massive central starbursts \citep{2010ApJ...715..202H}, however internal disk instabilities, either secular or merger-induced, are also expected to drive efficient radial inflows that can lead to a build up of central stellar mass density \citep{2007ApJ...670..237B,2014MNRAS.438.1870D,2015MNRAS.451.2933B}.  %{\color{BrickRed} How does kinematics change with the same processes?}
  %However, while this transition from late- to early-type morphology is often treated as a one-way street, particularly at low-redshift, the concept of morphology in hierarchical formation models is implicitly fluid (e.g. Brennan et al. 2015): galaxies move back and forth between the spheroid- and disk-dominated populations over their lifetimes, depending on the relative balance of mergers and smooth accretion. 

The same mechanisms are expected as possible pathways from star-forming to quiescence in the early universe. In high-redshift disks violent internal instabilities and gas-rich mergers could lead to dissipative processes able to form a massive, compact bulge that can either exhaust the available gas or stabilize it against further collapse \citep{2009ApJ...707..250M,2014MNRAS.438.1870D,2014ApJ...785...75G,2015MNRAS.450.2327Z}.  In such a scenario, star formation is expected to wind down on $\sim$Gyr timescales, resulting in young, quenched spheroids \citep{2014MNRAS.438.1870D,2014ApJ...791...52B}.  

%
%{\color{BrickRed} compact SFGs have received much attention as possible transition population}
Some models have shown that prior to being quenched galaxies exist in a compact-core or dense phase for a short time while still hosting significant star formation \citep{2014MNRAS.438.1870D,2015MNRAS.449..361W,2016MNRAS.458..242T}. These galaxies, while rare, have been increasingly identified in large data sets that have become available from both observations (e.g.~CANDELS; \citealt{2011ApJS..197...35G,2011ApJS..197...36K}) and simulations (e.g. EAGLE: \citealt{2015MNRAS.446..521S,2015MNRAS.450.1937C}; ILLUSTRIS: \citealt{2014MNRAS.444.1518V,2014MNRAS.445..175G}). Thus, these `compact star-forming galaxies' are possible immediate progenitors to distant compact quiescent galaxies. Observationally they are dense and compact ($<3$ kpc; \citealt{2013ApJ...765..104B,2014ApJ...791...52B,2014Natur.513..394N}), dusty \citep{2015ApJ...813...23V}, and despite often being branded as `blue nuggets', to have red colors \citep{2013ApJ...765..104B}. While such characteristics are in line with expectations from `wet compaction', there is also observational support for the scenario in which compact SFGs grow inside-out from already compact higher redshift progenitors \citep{2015ApJ...813...23V,2015MNRAS.449..361W}.
%However, these small star-forming systems may also be the tail-end of normal size evolution occurring at these redshifts from inside-out growth \citep{2015ApJ...813...23V}, as supported by some pathways in the Illustris simulations \citep{2015MNRAS.449..361W}.

While the sizes, colors, and star formation rates (SFRs) of these possible progenitors are well characterized, the dynamics of these systems are still being pieced together. Due to their small sizes, obtaining resolved dynamical information has been difficult \citep{2014ApJ...795..145B,2014Natur.513..394N,2015ApJ...813...23V}. Given their high central masses it is predicted that these objects will also have high central velocity dispersions both in gas and stars \citep{2012ApJ...753..167B,2012ApJ...751L..44W,2013ApJ...776...63F,2015ApJ...799..148B, 2015ApJ...813...23V, 2017ApJ...846..120B}, further linking them to quiescent galaxies at similar redshifts \citep{2013ApJ...771...85V,2014ApJ...783..117B}. Indeed, the first spectral measurements of compact SFGs at $z\sim2$ have revealed large integrated gas velocity dispersions, $\sim200$ \kms~\citep{2014ApJ...795..145B,2014Natur.513..394N}, comparable to stellar velocity dispersions of massive quiescent galaxies at similar redshifts. However, to what extent rotation, dispersion, and galactic winds contribute to the integrated linewidths is unknown, thus making use of the integrated linewidth as a dynamical mass indicator highly uncertain.
%How to interpret the unresolved kinematics is unclear given that compact star-forming galaxies share key physical properties with both star-forming and quiescent galaxies.

The past 15 years have revealed that massive quiescent galaxies in the local Universe feature a mixture of kinematic signatures, defined by both rotation and pressure support (e.g. \citealt{2007MNRAS.379..401E,2011MNRAS.414..888E,Cappellari:2011uq}). Given these results and the prevalence of massive high-redshift disk galaxies with prominent bulges (e.g. \citealt{2014ApJ...788...11L}) it is highly likely that at least a fraction of high-redshift quiescent galaxies and their progenitors also show disk-like morphologies and/or axial ratio distributions. Imaging studies lend credence to this hypothesis with 65\% of compact quiescent galaxies showing disk-like morphologies \citep{2011ApJ...730...38V,2013ApJ...773..149C}. Furthermore, stellar rotation $>100$ \kms~has now been measured directly in 2 fortuitously strongly lensed $z>2$ quiescent galaxies \citep{2015ApJ...813L...7N,2017Natur.546..510T}. Recent long- and multi-slit results of possible progenitors to compact quiescent galaxies have also shown evidence for rotation-dominated kinematics \citep{2015ApJ...813...23V}.% Furthermore, (they) are predicted to have low angular momentum... etc.

An additional factor that may contribute to the shut-down of star-formation in these systems is the role of active galactic nuclei (AGN) and AGN feedback \citep{2006MNRAS.365...11C,2006MNRAS.370..645B}. Compact SFGs are preferentially massive, thus high AGN fractions are expected (e.g. \citealt{2003MNRAS.346.1055K,2005ApJ...633..748R,2009AA...507.1277B}). 
This is reflected in modern simulations where the black hole accretion rate is tied to the the density of the surrounding gas \citep{2015MNRAS.449..361W}. The observational signatures of AGN and AGN-driven outflows are commonly seen in the kinematics of high-redshift galaxies and increasingly prevalent at the massive end of the star-forming `main sequence' (e.g. \citealt{Shapiro:2009sj,Alexander:2010bd,2014ApJ...787...38F,2014ApJ...796....7G}, F\"orster Schreiber et al.~\textit{in prep}).

Here we exploit the 3D information and depth of our \kmostd survey (\citealt{2015ApJ...799..209W}; hereafter W15), to take the next step in addressing the kinematic nature of compact SFGs, set quantitative constraints on the processes driving their emission line widths, and shed new light on the connection to compact quiescent galaxies at similar redshifts. 

We present the first integral field spectroscopic observations of 35 compact dense-core SFGs. In Section~\ref{sec.data} we discuss selection techniques of compact and dense galaxies. In Section~\ref{sec.results} we present our resolved \kmostd results and investigate kinematic tracers of the potential using the rotational velocity and integrated linewidths. In Section~\ref{sec.dis} we discuss the implications of our results on the possible evolutionary pathways for both creating and evolving compact SFGs. We conclude our results in Section~\ref{sec.conc}. We assume a $\Lambda$CDM cosmology with $H_0 = 70$ km s$^{-1}$ Mpc$^{-1}$, $\Omega_m = 0.3$, and $\Omega_\Lambda = 0.7$. For this cosmology, $1''$ corresponds to $\sim7.8$ kpc at $z = 0.9$, $\sim8.2$ kpc at $z = 2.3$ and $\sim7.2$ kpc at $z=3.6$. We adopt a \cite{2003PASP..115..763C} initial mass function.

\begin{figure*}[t]
\begin{center}
\includegraphics[angle=90, scale=0.7, trim=9cm 0cm 1cm 0cm]{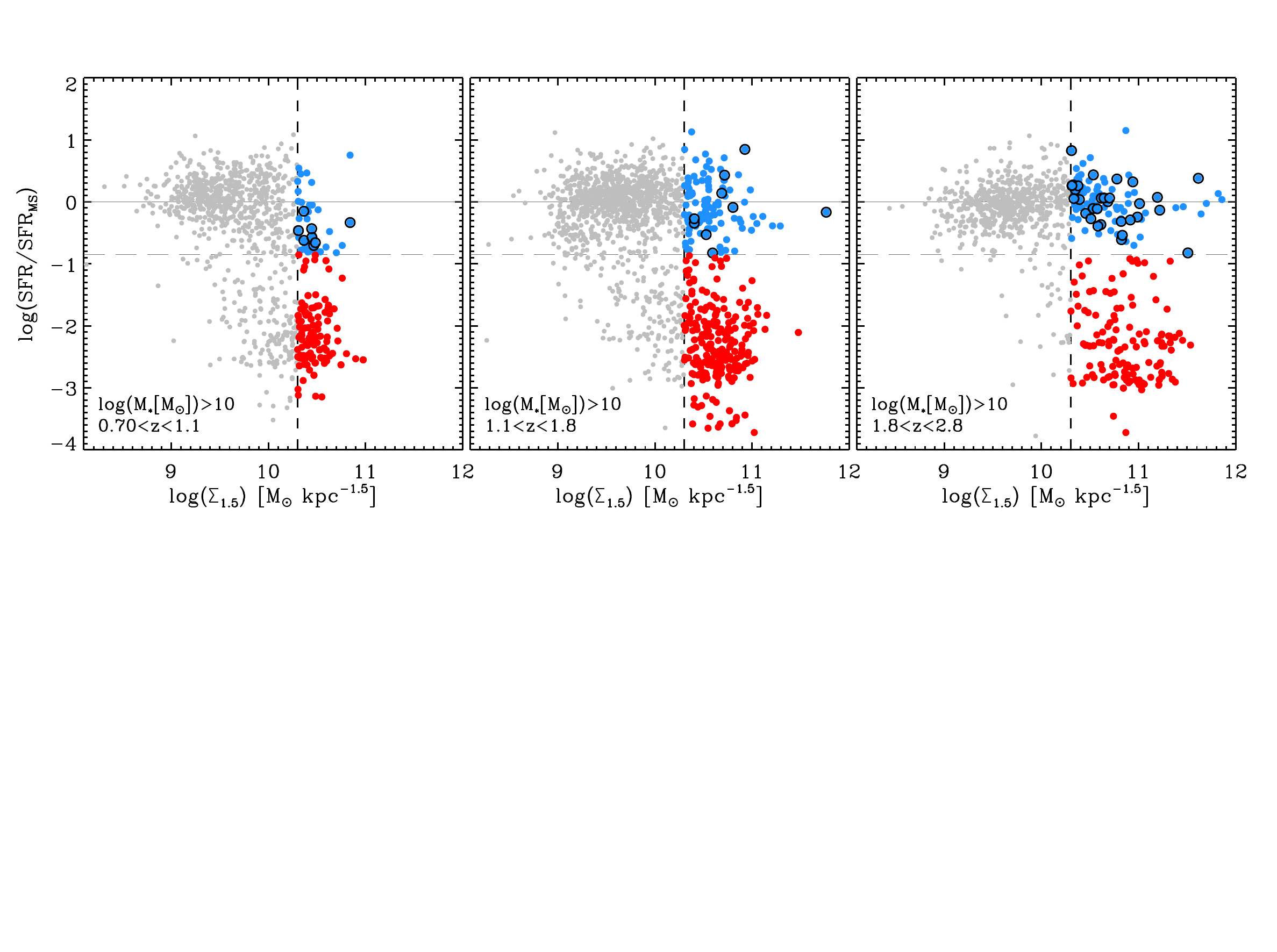}
\includegraphics[angle=90, scale=0.7, trim=8cm 0cm 0cm 0cm]{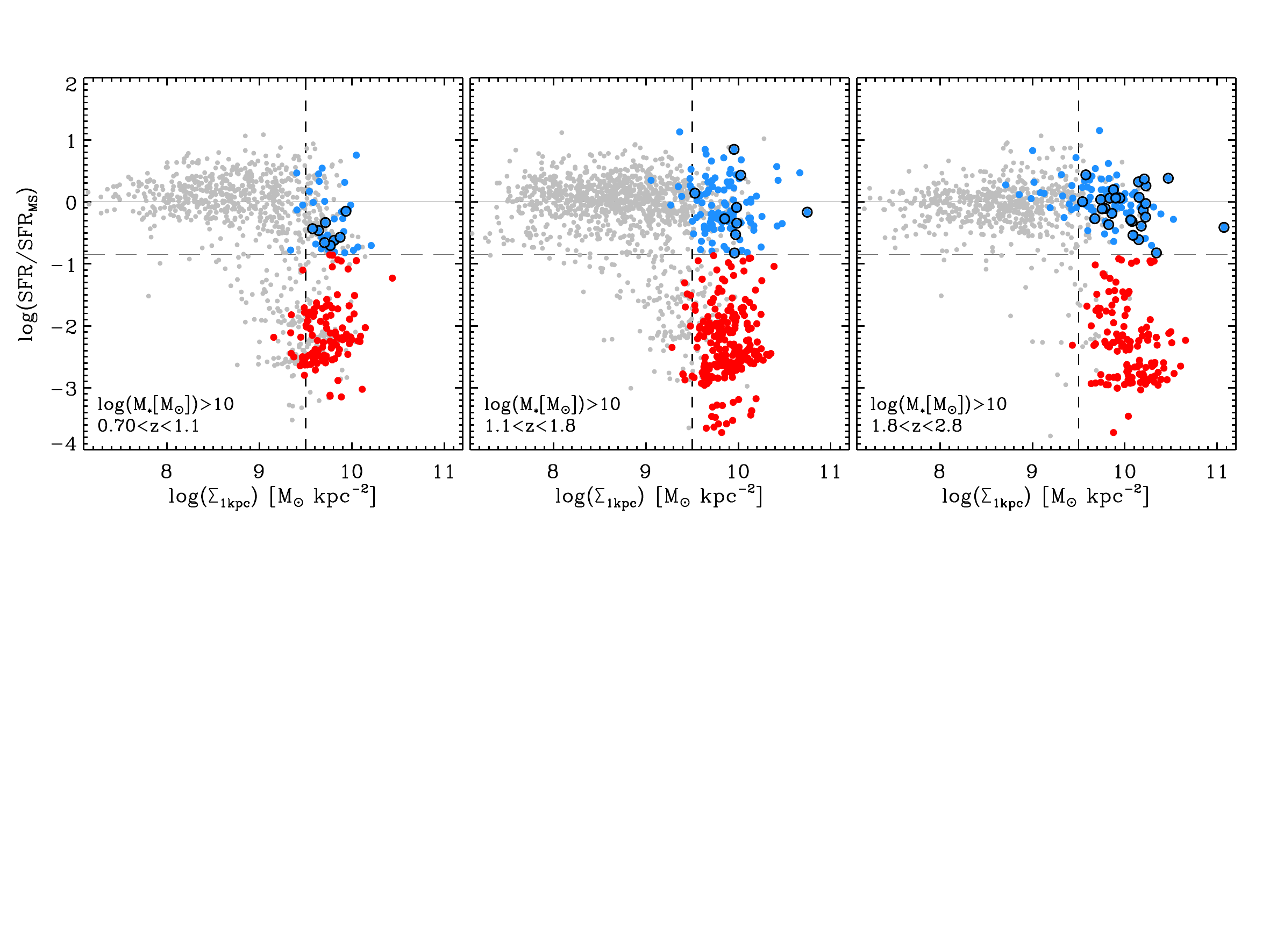}
\caption{SFR/SFR$_\mathrm{MS}$ vs. compactness, $\Sigma_{1.5}$ (top panels), and inner-kpc density, $\Sigma_\mathrm{1kpc}$ (bottom panels), as measured from HST $H$-band half light sizes for galaxies in v4.1.5 3D-HST/CANDELS satisfying the \kmostd magnitude and visibility selection criteria. The three columns show the selection diagram in three redshift bins, $0.7<z<1.1$, $1.1<z<1.8$, and $1.8<z<2.8$. The horizontal solid line denotes the canonical main sequence as defined by the 3D-HST main sequence (MS; \citealt{2014ApJ...795..104W}). Blue circles are SFGs (\dsfr$>-0.85$; horizontal dashed line) with $\log{(\Sigma_{1.5})}>10.3$ (vertical dashed line) and red circles are quiescent galaxies (\dsfr$<-0.85$) with $\log{(\Sigma_{1.5})}>10.3$. Black open circles highlight galaxies observed in \kmostd that have been selected as compact SFGs following the criteria $\log({\Sigma_{1.5}})>10.3$, $\log({\Sigma_\mathrm{1kpc}})>9.5$, \dsfr$>-0.85$, and $M_\ast>10^{10}$~\Msun.}
\label{fig.select}
\end{center}
\end{figure*}

\section{Data}
\label{sec.data}
%%%%%%%%%%%%%%%%%%%%%%%
\subsection{KMOS observations and data analysis}
\label{sec.obs}
%%%%%%%%%%%%%%%%%%%%%%

\kmostd is an ongoing kinematic survey using the K-band Multi-Object Spectrograph \citep[KMOS;][]{2013Msngr.151...21S} to obtain near-infrared (IR) integral field spectroscopy covering the \NII+\halpha emission line complex at $0.7 < z < 2.7$.  Targets were selected from the 3D-HST Legacy Survey \citep{2014ApJS..214...24S,brammer:2012:04,2016ApJS..225...27M} and CANDELS \citep{2011ApJS..197...35G,2011ApJS..197...36K} in COSMOS, GOODS-S, and UDS, and the primary sample includes galaxies with $K_s \leq 23$ and a prior redshift (3D-HST) that places \halpha in a region of the spectrum relatively free of contamination from OH sky lines.  In addition, we also include data covering \OIII+\hbeta for narrow-band selected galaxies extending up to $z = 3.7$ also observed in \kmostd pointings.

The \kmostd sample includes galaxies spanning more than 4 orders of magnitude in specific star formation rate (sSFR) because the target selection does not include any \textit{a priori} information on SFR. In this work we identify SFGs based on their positions relative to the main sequence such that $\Delta\mathrm{SFR}\equiv\log{(\mathrm{SFR_\mathrm{UV+IR}/SFR}_\mathrm{MS})}>-0.85$, where SFR$_\mathrm{MS}$ is defined by the 3D-HST main sequence (MS; \citealt{2014ApJ...795..104W}) for each galaxy given its redshift and stellar mass using the parameterization of W15.  SFRs are derived combining the unobscured (UV) and obscured (IR) star formation following \cite{2011ApJ...738..106W} or from the spectral energy distribution (SED) if there is no IR detection.  We note that the adopted cut between star-forming and passive galaxies for this paper is similar to an evolving $UVJ$ selection; however, for the purposes of investigating possible progenitors of quenched galaxies we do not want to rule-out galaxies with $UVJ$ passive colors that have residual star-formation \citep{2013ApJ...765..104B,2017ApJ...841L...6B}. 

%Observational details
Full details of the \kmostd observing conditions, observing strategy, and data reduction procedure are given in W15 and a forthcoming data release paper. Relevant details for this work  are described briefly in the following paragraphs.

Observations of \kmostd data used here have been carried out between October 2013 and September 2017 with exposure times ranging from 3--30 hours on source. Data was collected in excellent seeing conditions, $YJ$, $H$, or $K$ band median seeing FWHM$=$ 0.55$''$. The model-independent seeing is measured as the FWHM from stars observed simultaneously in the same waveband and same detector as the galaxy observations. For the compact sources discussed in greater detail in this paper the median seeing was 0.5$''$ with individual values ranging from $0.42-0.61$.

We map the kinematics across the emission-line detected regions of the galaxy using single Gaussian fits after applying a $3\times3$ pixel spatial median filter. The median filtering is used only to create the kinematic map. Integrated spectra and 1D kinematic extractions are extracted directly from the original data cube. The kinematic axis is determined from the 2D velocity field as the direction of the largest observed velocity difference. One-dimensional kinematic profiles are extracted along the kinematic major axis using an aperture equivalent to the PSF, unique to each galaxy. When a kinematic axis cannot be determined from the 2D velocity field the photometric major axis is used for the 1D extractions. Galaxies are considered resolved when emission lines can be fit to a radius of 1.5$\times$ the PSF FWHM from the center of the continuum emission. The 1D kinematic profiles allow measurements slightly beyond the extent of individual spaxels as they are measured from summed spectra within an aperture. In resolved cases we derive an estimate of the observed velocity difference along the kinematic major axis, $v_\mathrm{obs}$, rotational velocity, $v_\mathrm{rot}=v_\mathrm{obs}/\sin{i}$, and disk velocity dispersion, $\sigma_0$. We estimate $\sin{i}$ using the HST $H$-band (F160W) structural axis ratio ($b/a$) assuming a thick disk with ratio of scale height to scale length of 0.25. Structural parameters are drawn from \cite{2014ApJ...788...11L} based on single component GALFIT \citep{2002AJ....124..266P} models (see also \citealt{2014ApJ...788...28V}). We derive integrated spectra for all \kmostd galaxies by summing over spaxels within a 1.5$"$ diameter aperture.  The velocity gradients are not removed from these spectra. Good spaxel masks are created for resolved galaxies. They include spaxels where the S/N of \halpha or \OIII~is $>5$, velocity uncertainties of $<100$ km s$^{-1}$, and/or a relative velocity dispersion uncertainty of $<50$\%. 
%We also measure the linewidth within one resolution element centered on the continuum image of the galaxy, $\sigma_\mathrm{cent}$.  

%Beam smearing
KMOS observations are seeing limited and subject to beam smearing. Beam smearing corrections used to calculate the corrected rotational velocity, $v_\mathrm{rot,corr}$, and disk velocity dispersion, $\sigma_\mathrm{0,corr}$, are derived following the methods in Appendix A.2.4 of \cite{2016ApJ...826..214B}. In short, multiplicative beam smearing corrections are derived from a set of exponential disk dynamical models run at different inclinations, masses, ratios of half-light size to PSF size, velocity dispersions, and instrumental resolutions. To determine a beam smearing correction for an individual galaxy a relation relating the ratios of half-light size / PSF size to the magnitude of the beam smearing correction is queried based on the properties of the specific galaxy and observed PSF. The typical magnitude of the beam smearing corrections for the compact sources discussed here is discussed at the end of Section~\ref{subsec.resolved}.
For a more detailed description of the methods used to derive the kinematic maps and subsequent parameters we refer the reader to W15 and Appendix A of \cite{2016ApJ...826..214B}.

\subsection{Compact galaxy selection}
\label{sec.sel}

Our selection of compact galaxies relies primarily on two parameters: the global compactness $\Sigma_{1.5}$ ($\equiv M_\ast/r_e^{1.5}$), where $r_e$ is the HST $H$-band (F160W) circularized half-light size \citep{2013ApJ...765..104B}, and the stellar surface density within the central 1 kpc, \skpc{}.  We compute \skpc{} by integrating over the deprojected luminosity density distribution as described by \cite{2014ApJ...791...45V}. We additionally require that compact galaxies are massive, $M_\ast>10^{10}$~\Msun. Figure~\ref{fig.select} shows the distribution of \kmostd galaxies in terms of their star-formation rate, compactness (top panels) and central density (bottom panels) in three redshift ranges corresponding to the KMOS \textit{YJ}, \textit{H}, and \textit{K} bands.

\begin{figure}[t]
%\begin{center}
\includegraphics[scale=0.35, trim=0cm 0cm 0cm -1.25cm, angle=90]{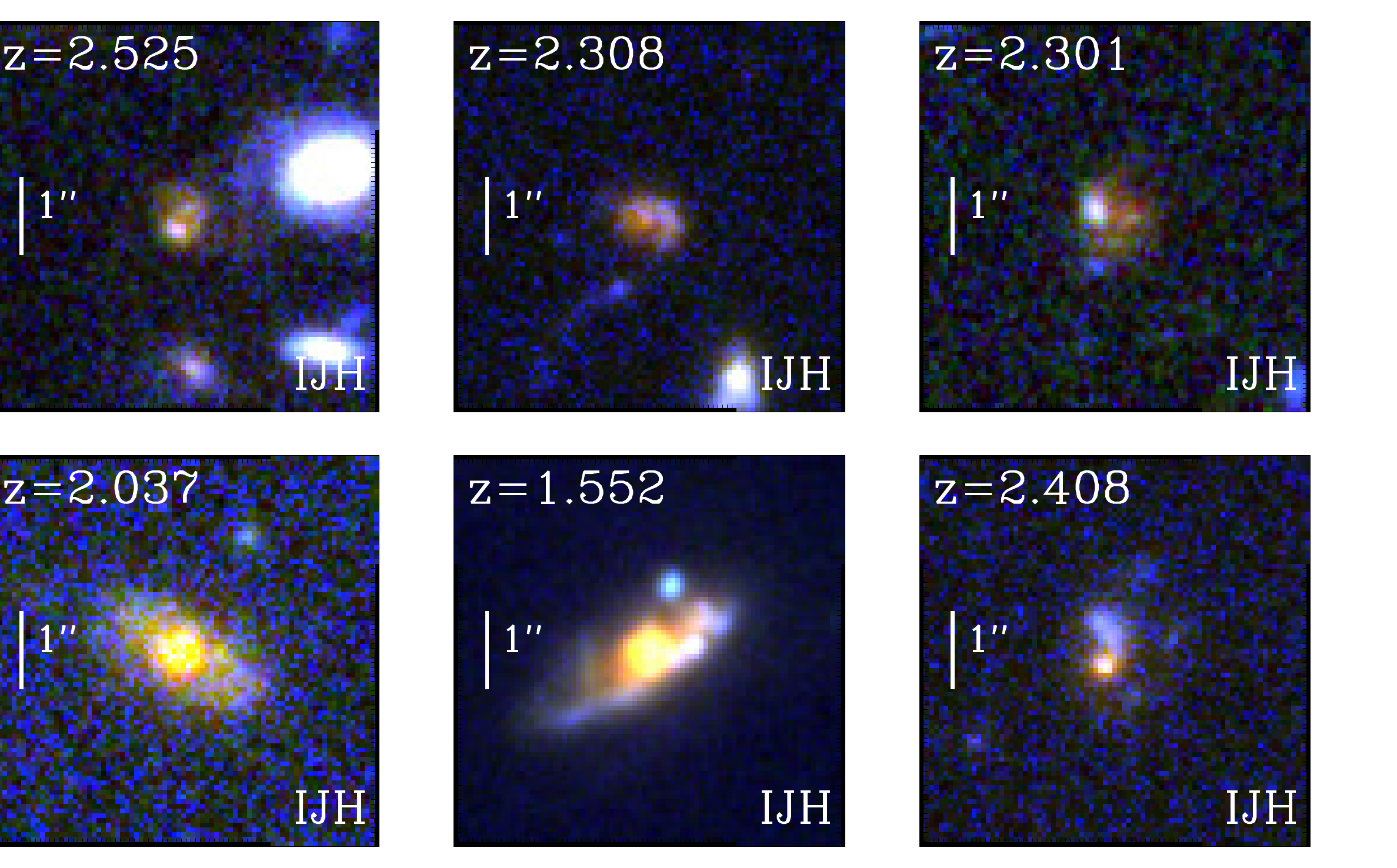}
\caption{Example composite HST $IJH$ images showing galaxies exclusively satisfying either the selection of compact galaxies ($\log({\Sigma_{1.5}[\mathrm{\Msun~kpc}^{-1.5})]}>10.3$; top) or of dense core galaxies ($\log{(\Sigma_\mathrm{1kpc}[\mathrm{\Msun~kpc}^{-2}])}>9.5$; bottom). }
\label{fig.morpho}
%\end{center}
\end{figure}

While selecting on global compactness identifies galaxies morphologically similar to compact quiescent galaxies (axis ratios of approximately unity and small sizes; $r_e<2$ kpc), selecting on central density identifies galaxies with a variety of sizes and axial ratios but with the presence of a dense stellar core, often seen as a requirement for quenching (e.g. \citealt{cheung:2012aa,2014ApJ...788...11L,2014MNRAS.441..599B,2014ApJ...791...45V}). In all panels of Figure~\ref{fig.select} the blue points show the SFGs selected on global compactness using $\log(\Sigma_{1.5}) > 10.3$ \citep{2013ApJ...765..104B}. Their location in the bottom panels shows the overlap between what would be selected as compact rather than as containing a dense core using $\log(\Sigma_\mathrm{1kpc}) > 9.5$. Galaxies observed in \kmostd are identified by black open circles in the top panels. Figure~\ref{fig.morpho} shows examples of the types of galaxies selected by these two different criteria separately.  While the galaxies in the top 3 panels are compact, they do not closely resemble quiescent galaxies at $z\approx1-3$, which is a key motivation to study the compact SFGs. In contrast, the bottom three panels show galaxies with bulge components reminiscent of quiescent galaxies surrounded by large blue disks. We adopt as our final `compact' sample those galaxies satisfying $both$ the global compactness and central density criteria ($\log({\Sigma_{1.5}})>10.3$ and $\log({\Sigma_\mathrm{1kpc}})>9.5$).

Of the {502} `star-forming' galaxies, $\Delta SFR>-0.85$, in our \kmostd sample (as of December 2016), {45} satisfy both the compactness and central density criteria, {35} of which are detected, spanning the redshift range $0.9<z<3.7$, including both \halphans(33) and \OIII (2) detected galaxies.  The composite HST $IJH$ images, integrated spectra, and exposure times for these objects are shown in Figure~\ref{fig.intspec1}, ordered by increasing redshift. Integrated spectra for each galaxy are created by summing the spaxels in a 1.5" diameter aperture. Single component Gaussian fits are over-plotted in red.  

The average half-light radius, \sersic index, axis ratio, and bulge-to-total (B/T) of the detected compact SFGs are {1.85 kpc, 3.9, 0.72, and 0.5 respectively}, compared to {1.36 kpc, 3.7, 0.65, and 0.7} for compact quiescent galaxies select using the same criteria. However, we note that some of these galaxies are close to the resolution limit of HST in the $H$-band resulting in large uncertainties in structural parameters not included in the formal errors. For example, of the 35 detected galaxies 6 have unconstrained \sersic indices of either 0.2 or 8.0, the limits of the fitting range. We note that 3 of the 6 are Type-I AGN as discussed in the next section. 

Ten of the detected compact SFGs have rest-frame colors consistent with being passive using a $UVJ$ color selection (e.g \citealt{2013ApJ...777...18M}). In addition to the compact SFG sample, we robustly detect \halpha in 15 compact quiescent galaxies (QG; $\Delta SFR<-0.85$), of which two are resolved. The properties of \halpha detected quiescent galaxies are discussed in \cite{2017ApJ...841L...6B}. 

%\subsubsection{Nuclear activity}
We find {76\%} of the detected compact SFGs may host an active galactic nucleus (AGN). In the general SFG population, AGN incidence is a strong function of stellar mass (e.g. \citealt{2003MNRAS.346.1055K}). In addition to optical indicators, this has also been shown using ultra-violet, infrared, and X-ray indicators (e.g. \citealt{2005MNRAS.362...25B,2008ApJ...681..931B}). 
Combining X-ray, IR, and radio data confirm that the high mass galaxies, and in particular the compact dense galaxies in \kmostd have high AGN fractions, as also shown by \cite{2013ApJ...765..104B} and \cite{2017ApJ...846..112K}. Using techniques described in \cite{2014ApJ...796....7G} to identify likely AGN hosts we recover this trend in our full observed sample.

\begin{figure*}[t]
\begin{center}
\includegraphics[scale=1.0, trim=1cm 0cm 0cm 0cm]{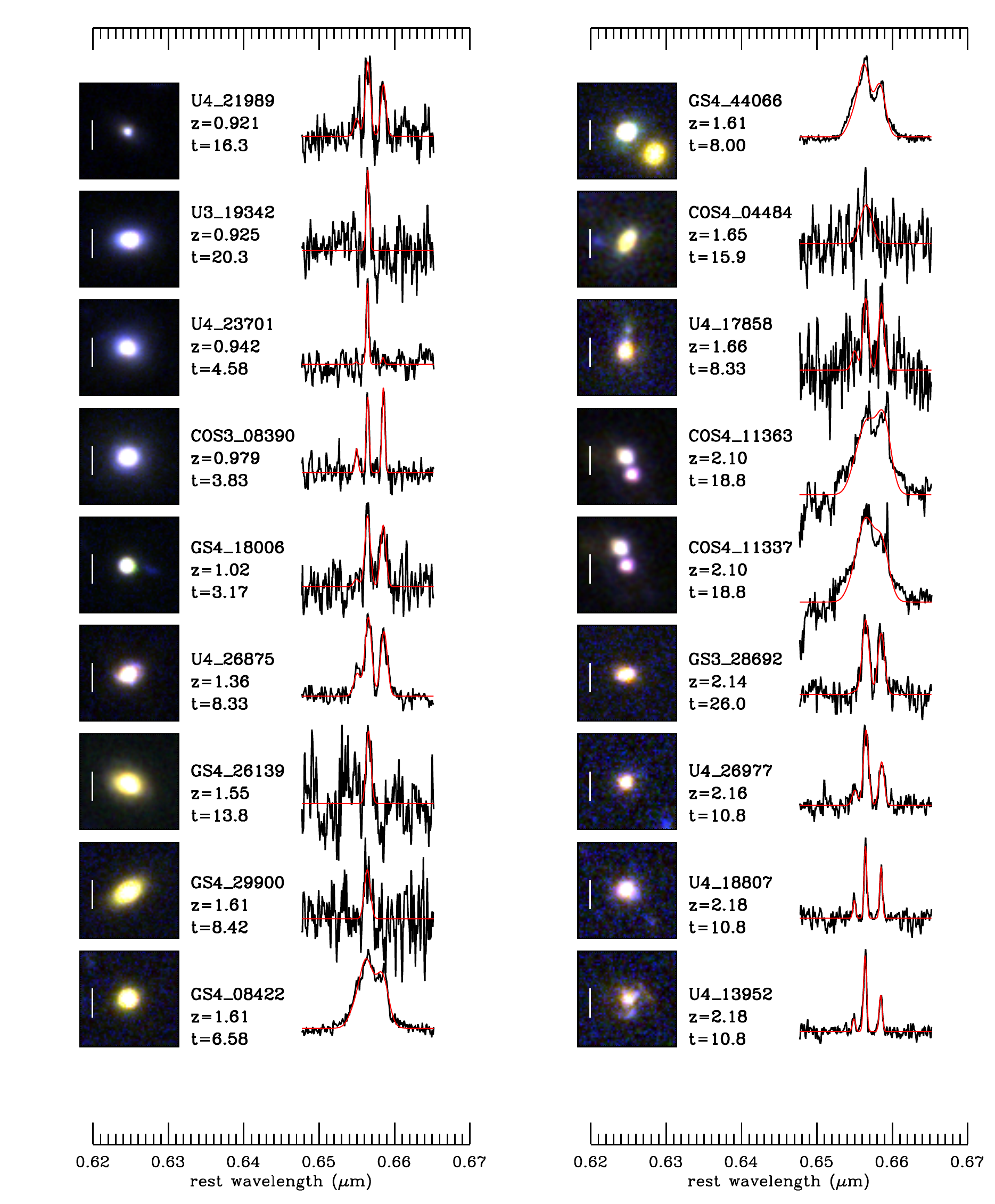}
\caption{HST observed-frame $IJH$ images and \kmostd integrated spectra of \halphans-\NII~or \hbeta-\OIII~emission complexes, for the detected galaxies that are both centrally dense and compact as defined by $\log{(\Sigma_{1.5})}>10.3$ and $\log{(\Sigma_\mathrm{1kpc})}>9.5$. The spectra are sigma-clipped and normalized to arbitrary flux units. Single Gaussian fits to the \halphans-\NII~or \hbetans-\OIII~complex are overlaid in red. Observation times, $t$, are given in hours for each object. White bars in the images show 1 arc sec.  Galaxies are shown in order of increasing redshift. We note that the spectra of COS4\_11363 and COS4\_11337 are blended in the \kmostd data due to their small separation on the sky.}
\label{fig.intspec1}
\end{center}
\end{figure*}

\addtocounter{figure}{-1}
\begin{figure}[t!hbp]
\begin{center}
\includegraphics[scale=1.0, trim=1cm 0cm 8cm 0cm, clip]{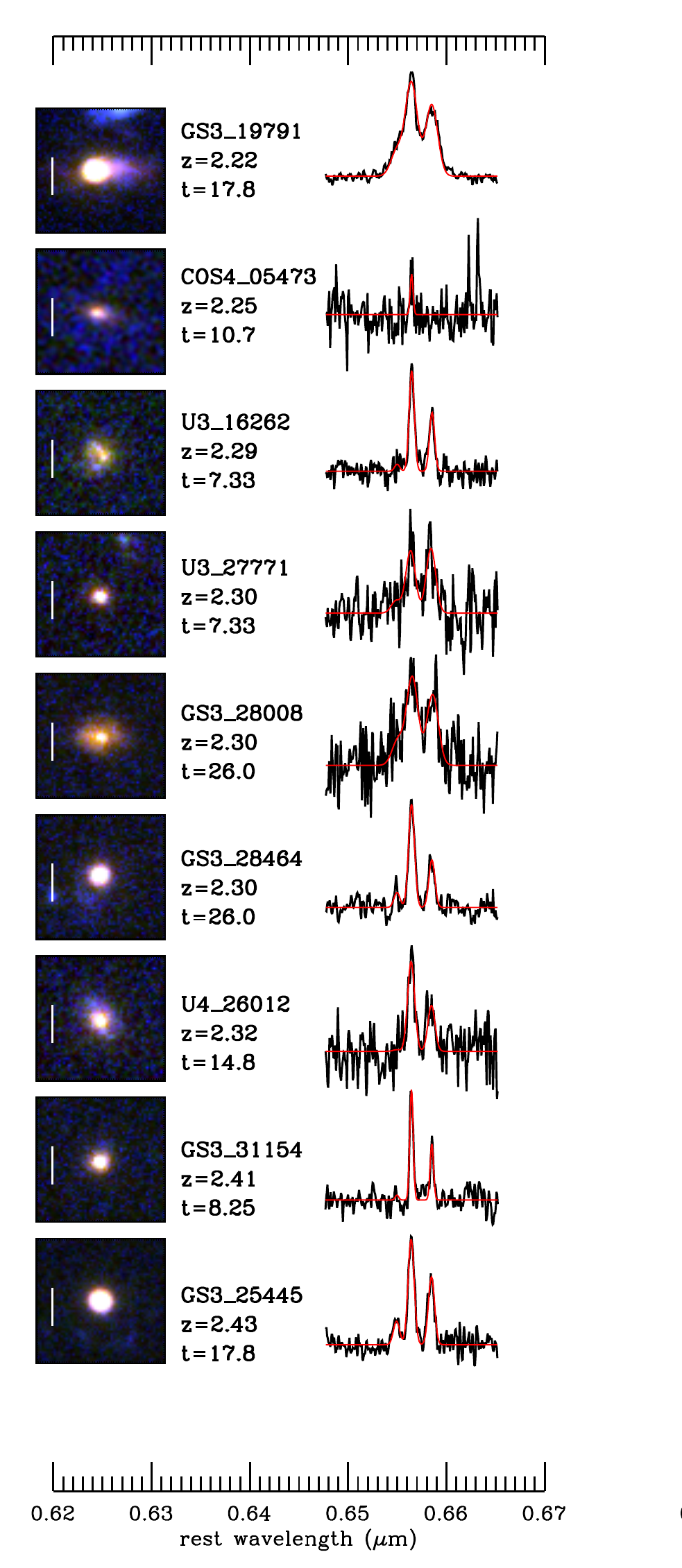}\\
\caption{Continued from previous page.}
\end{center}
\end{figure}
\begin{figure}[thbp]
\begin{center}
\includegraphics[scale=1.0, trim=10.6cm 9cm 0cm 0cm, clip]{fig3b.pdf}
\includegraphics[scale=1.0, trim=10.6cm 0.3cm 0cm 21cm, clip]{fig3b.pdf}
\includegraphics[scale=1.0, trim=1cm 17.5cm 0cm 0.17cm, clip]{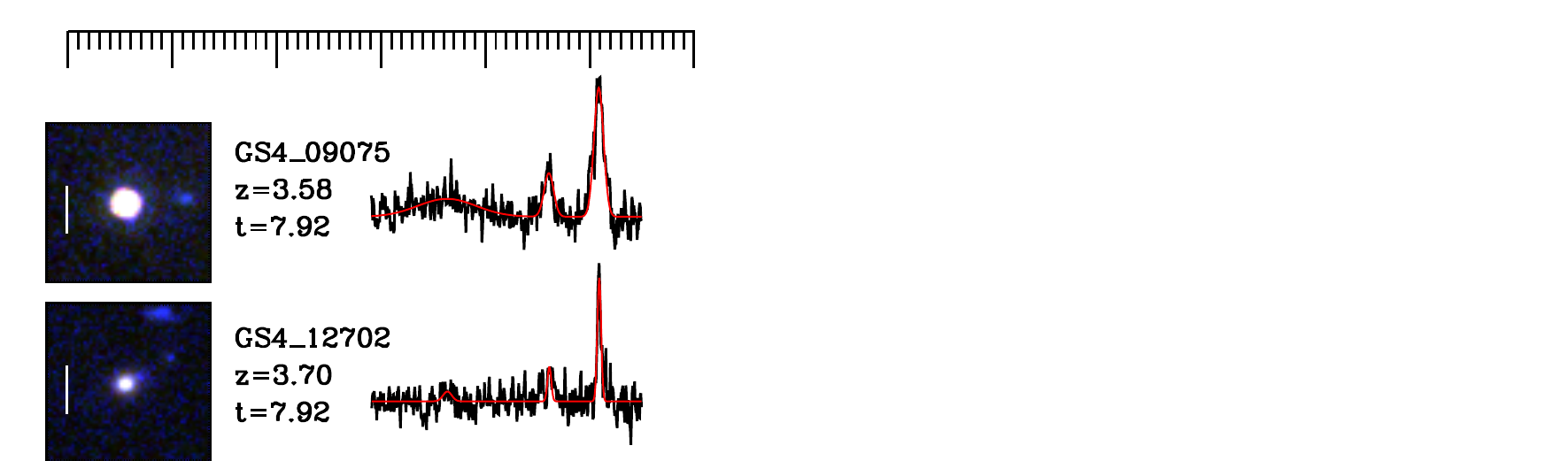}
\includegraphics[scale=1.0, trim=1cm 0cm 0cm 21cm, clip]{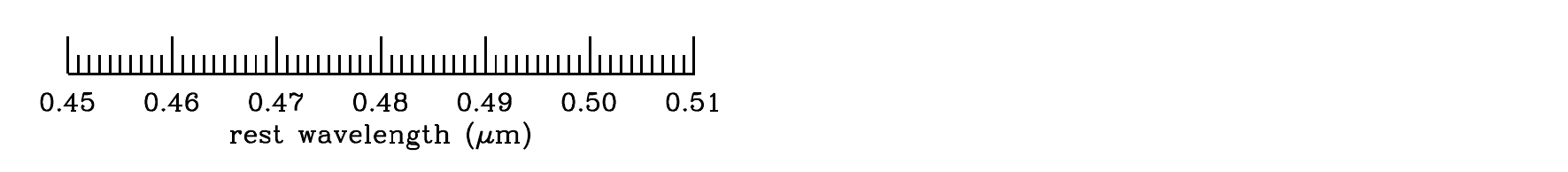}
\end{center}
\end{figure}

In concordance, nuclear activity is prevalent in the \kmostd high-mass galaxies as determined from high \NII/\halpha ratios ($\log$(\NII/\halphans)$ >-0.1$) and the presence of an underlying broad \halpha emission component (\citealt{2014ApJ...796....7G}, F\"orster Schreiber et al.~\textit{in prep}). These results are consistent with the rapid increase of outflow incidence  as a function of stellar mass as shown in our previous studies \citep{2014ApJ...787...38F,2014ApJ...796....7G} and updated with approximately 5$\times$ larger sample (F\"orster Schreiber et al.~\textit{in prep}), as well as with general trends with increasing central stellar mass density. We find that 45\% of all observed \kmostd SFGs exhibit high \NII/\halpha or a broad component in stellar mass range $10.5<\log{(M_*[\mathrm{\Msun}])}<11.7$. Dense compact galaxies are among the most massive \kmostd galaxies, by selection, and show $\sim1.2\times$ higher fractions of spectral signatures of nuclear activity than the full population, with  55\% in the same mass range. When taking into account all five AGN indicators from \cite{2014ApJ...796....7G} for the full \kmostd sample in stellar mass range $10.5<\log{(M_*[\mathrm{\Msun}])}<11.7$ the AGN fraction is 53\% compared to 76\% for dense compact SFGs.  

Fitting a single Gaussian to the spectrum of a galaxy hosting a  outflow can result in an artificially larger linewidth with the strongest effect most likely to occur for shallow data strongly light-weighted by bright central regions where the presence of a centrally driven wind can dominate \citep{2014ApJ...787...38F,2014ApJ...796....7G}. This effect is investigated further in Section~\ref{sec.linewidths}.

Three compact SFGs are possible Type-I AGN as indicated from the combination of KMOS spectra, X-ray, and rest UV data, two of which have $\sigma_\mathrm{tot}>500$ \kms~from a single Gaussian fit to the KMOS data.  The HST images of these three galaxies, COS4\_14596, COS4\_21492, and GS4\_09075, exhibit characteristics of the PSF such as diffraction spikes and spots. We exclude these objects from our further analysis of the integrated linewidth and emission line ratios.

%%%%%%%%%%%%%%%%%%%%%%%%%%%%%%%%%%%%%%%%%%%%%%%%
\section{Results}
\label{sec.results}
%%%%%%%%%%%%%%%%%%%%%%%%%%%%%%%%%%%%%%%%%%%%%%%%
\subsection{\halpha sizes}
\label{sec.sizes}
We measured intrinsic \halpha or \OIII~sizes for galaxies in our compact SFG sample by fitting a pure exponential disk profile convolved with the KMOS PSF to two-dimensional KMOS emission maps, with the centroid, ellipticity, and PA fixed to that of the continuum map in the HST F160W band \citep{2015ApJ...799..226E}.  The flux in individual spaxels was estimated by integrating the continuum-subtracted spectrum within a $\pm200$ km s$^{-1}$ window centered on the expected wavelength of the emission line, derived either from a single-component Gaussian fit to the line or, in the case of low-S/N regions, the nearest spaxel with a successful fit.  In this way, we were able to derive emission maps extending over the full KMOS field of view.  For each galaxy we construct a model of the KMOS PSF by stacking registered images of individual PSF stars taken in the same exposure and detector.  We note that, due to uncertainties in the relative position of KMOS arms ($\sim$0.2 arcsec r.m.s) this procedure likely underestimates the "true" PSF FWHM, however we do not expect it to significantly affect our conclusions.  Uncertainties on the sizes were estimated from bootstrap realizations of the combined KMOS data, where individual 300s exposures were randomly recombined with replacement.  Further details of the modeling procedure will be described in a future paper (Wilman et al.~\textit{in prep.}).

The average intrinsic \halpha half-light size of the compact SFGs is $2.5\pm0.2$ kpc. The \halpha sizes are typically between $1-2\times$ the continuum sizes as measured from single S\'ersic fits to the CANDELS F160W images, with 33\% agreeing with a size ratio of unity within 1$\sigma$ errors.  
The average size ratio, $r_{\mathrm{H}\alpha}/r_e\mathrm{[F160W]}$, is 1.2 but ranges between 0.7 and 4.1. The size ratios are consistent with size ratios found for the most massive galaxies in the 3D-HST survey at $z\sim1$ \citep{2016ApJ...828...27N} and the \halpha sizes of the \kmostd compact SFGs are comparable to those derived from position-velocity diagrams of long- and multi-slit observations of compact SFGs \cite{2015ApJ...813...23V}.

\halpha sizes also show good agreement with the rest-frame $UV$ emission as probed by the observed $I$-band distribution at these redshifts. Some compact SFGs in our sample do exhibit faint emission in the $I$- and $J$-bands as seen in the composite $IJH$ images of Figure~\ref{fig.intspec1} with either asymmetric emission around the dominant $H$-band light (e.g. U4\_17858 and COS4\_22995) or extended emission reminiscent of a faint disk or spiral features (e.g. GS3\_19791 and U4\_26012). The galaxies with visible features in the $I$-band are among the galaxies with the largest \halpha sizes. In contrast, some simulations predict faint outer disks or rings surrounding compact SFGs \citep{2015MNRAS.450.2327Z}. Galaxies in the \kmostd sample which do fit the description of centrally dense cores surrounded by large ($>2$ kpc) star-forming rings are the extended centrally dense galaxies shown in the bottom panels of Figure~\ref{fig.morpho}.

\subsection{Compact SFG un-resolved kinematics}
\label{sec.unresolved}

\begin{figure}[!t]
\begin{center}
\includegraphics[scale=0.75, angle=90, trim=10.7cm 12cm 0.75cm 0.5cm, clip]{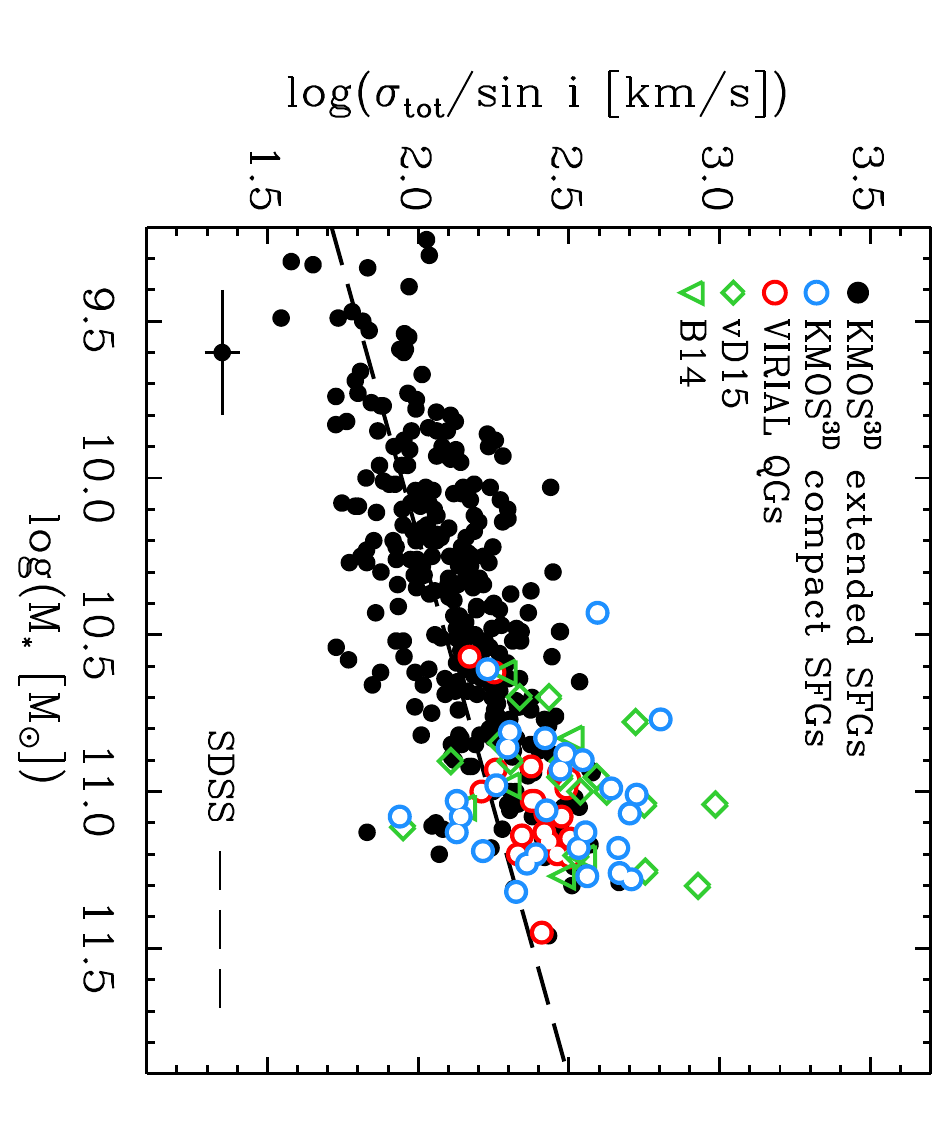}
\caption{Inclination corrected integrated linewidth vs. $M_*$.
%and $\Sigma_\mathrm{1kpc}$. 
The dashed line shows the relation derived from SDSS \citep{2013ApJ...776...63F}.  All SFGs from \kmostd are shown by black and blue circles, literature compact SFGs \citep{2014ApJ...795..145B,2015ApJ...813...23V} are shown by green diamonds, and quiescent galaxies from VIRIAL are shown by red circles. For star-forming galaxies \stot~is measured from \halpha or \OIII~emission. In the case of galaxies observed in \kmostd and the literature only the \kmostd galaxies are shown. For quiescent galaxies \stot~is measured from stellar absorption features (Mendel et al.~\textit{in prep.}). All star-forming galaxies with the exception of values from \cite{2014ApJ...795..145B} are corrected for inclination. A representative error bar is shown in the bottom left corner.
%The dotted lines show the best fit linear regressions to the \kmostd data. ...
}
\label{fig.trends}
\end{center}
\end{figure}

For all detected compact SFGs we measure an integrated linewidth, $\sigma_\mathrm{tot}$, from single Gaussian fits to the \halphans-\NII~or \OIII~ complex of the non-velocity shifted integrated spectrum and correct for spectral resolution. The linewidths of compact SFGs cover a wide range, $75-400$ \kms. In Figure~\ref{fig.trends} we show the relationship between \halpha velocity dispersion and stellar mass for extended and compact SFGs from \kmostdns. Compact SFGs from the literature are shown by green diamonds \citep{2015ApJ...813...23V,2014ApJ...795..145B}. We also compare our results to a complementary quiescent galaxy field survey with KMOS, the VLT IR IFU Absorption Line survey (VIRIAL; \citealt{2015ApJ...804L...4M}; \textit{in prep}). VIRIAL targets are are $UVJ$ passive galaxies selected from the 3D-HST survey. The velocity dispersions for the VIRIAL survey are measured from absorption lines of unresolved compact galaxies. Under the assumption that the majority of our sample is rotating we apply a simple $\sin{i}$ correction to all \kmostd galaxies to account for inclination effects (In W15 we show that 83\% of SFGs at these epochs are rotationally-supported). In contrast, we do not apply a $\sin{i}$ correction to the passive galaxy sample which we restrict to systems with $n>2$ \citep{2017ApJ...834...18B}.

Figure~\ref{fig.trends} shows that the compact SFGs form the high-mass end of the trend established from the general SFG population. The linewidths of compact SFGs are consistent with the linewidths of extended SFGs from the KMOS$^{\mathrm{3D}}$ survey of similar mass and redshift. Compact SFGs from the literature overlap but on average have slightly larger linewidths for their mass than when compared to the \kmostd data. This may be a result of selection, shallower data, or contamination by nuclear-driven outflows. 
In general, the \halpha velocity dispersions of compact SFGs show excellent agreement with the central stellar velocity dispersion of quiescent galaxies at equivalent redshifts and masses. We note that there are some ambiguities in comparing stellar and gas velocity dispersions, however because \stot~encompasses both rotation and turbulence it should trace the total dynamical mass of the galaxy comparable to central stellar dispersions (e.g \citealt{2006ApJ...653.1027W,2007ApJ...660L..35K,2013MNRAS.432.1862C,2014RvMP...86...47C,2017arXiv171007694G}). We investigate the relationship between \stot~and rotational and dispersion velocities in Section~\ref{sec.linewidths}.

%%Connection to local relations
Locally a tight relationship between central velocity dispersion and stellar mass has been established for both quiescent and star-forming galaxies \citep{2012ApJ...751L..44W,2013ApJ...776...63F}. We recover a correlation over two orders of magnitude between \stot~and $M_*$ with a consistent slope and factor of {$\sim1.38$} offset from the local relation parameterized by \cite{2013ApJ...776...63F} with 
SDSS\footnote{The local relation is established using fiber-based central stellar velocity dispersions corrected to a radius of 1 kpc using equation 1 of \cite{2006MNRAS.366.1126C}. We do not know if the intrinsic dispersion curves mirror those in \cite{2006MNRAS.366.1126C} and observational evidence indicates that they may be flat \citep{2009ApJ...697..115C} resulting in a null correction. However, if we assume that the dispersion curves have comparable radial profiles to the galaxies in \cite{2006MNRAS.366.1126C} and apply the same correction then the \kmostd linewidths would increase by $\sim$1.09, with the highest factors being 1.15 for the few galaxies with half-light radii $>8$ kpc and the smallest correction factors for compact SFGs and quiescent galaxies.}. 
The slight offset of the population from the local relation could be explained by evolutionary effects. For example, it is well known that for SFGs the integrated linewidth is correlated with rotational velocity, both locally and at high-redshift (e.g. \citealt{1985ApJS...58...67T,2006ApJ...653.1027W}). Thus Figure~\ref{fig.trends} roughly presents a stellar Tully-Fisher Relation \citep{1977A&A....54..661T,2017ApJ...842..121U}. The details of the relationship between \stot~and velocity are less clear and may also change with redshift. This is explored in greater detail for the full \kmostd sample in Section~\ref{sec.linewidths}.

%%%%%%%%
\subsection{Compact SFG resolved kinematics}
\label{subsec.resolved}
\begin{figure*}[thbp]
\begin{tabular}{ccc}
\vspace*{-1.5em}
\hspace*{-2em}\subfloat{\includegraphics[scale=0.438,  trim=0.0cm 3.0cm 0.0cm 0cm, clip, angle=90]{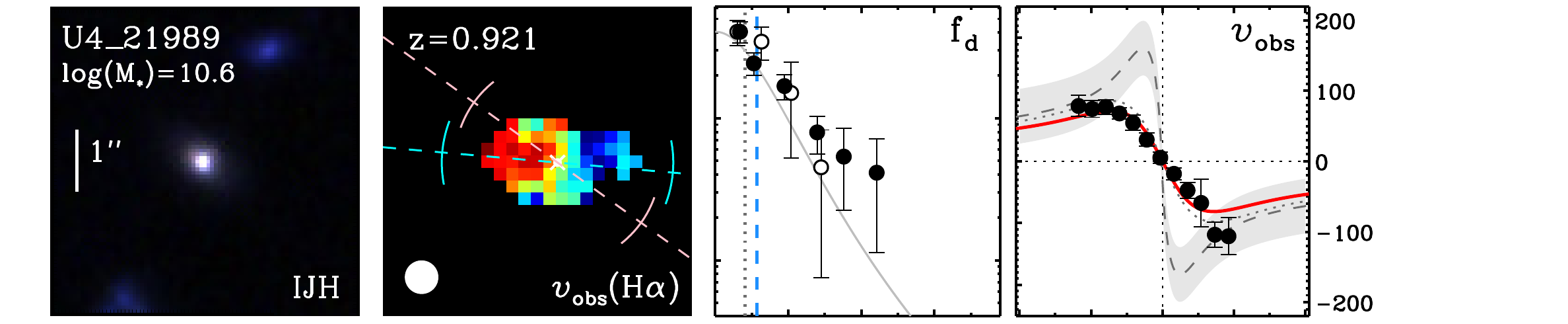}} &\hspace*{-2em}
\subfloat{\includegraphics[scale=0.438,  trim=0.0cm 3.0cm 0.0cm 0cm, clip, angle=90]{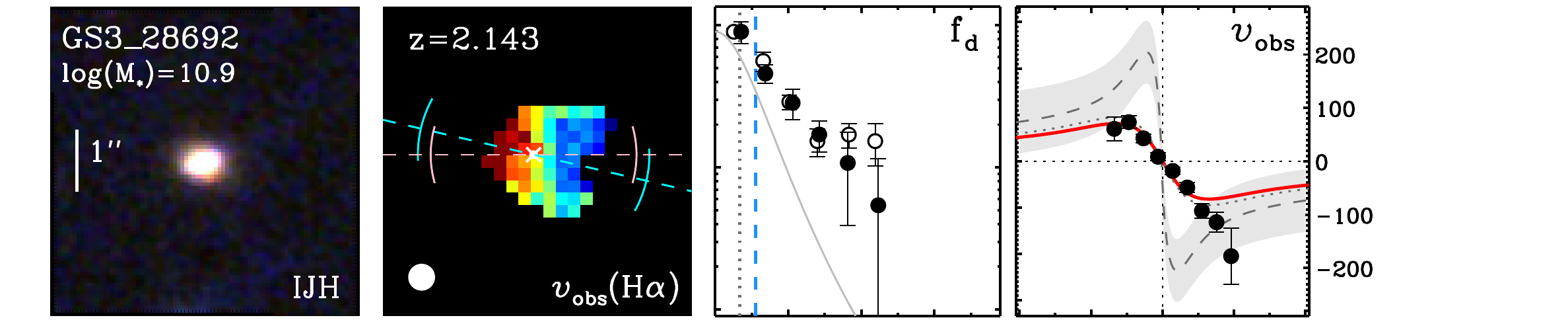}}\\
\vspace*{-1.5em}
\hspace*{-2em}\subfloat{\includegraphics[scale=0.438,  trim=0.0cm 3.0cm 0.0cm 0cm, clip, angle=90]{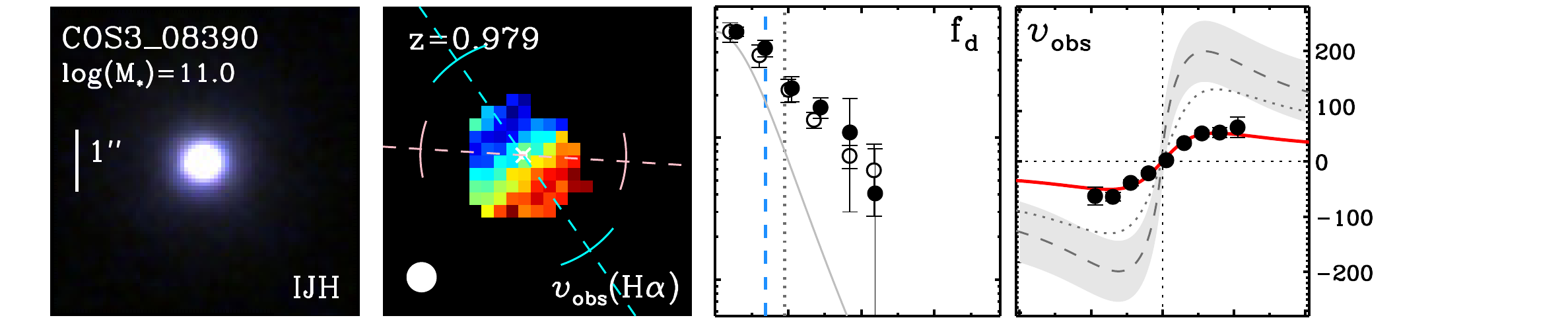}} &\hspace*{-2em}
\subfloat{\includegraphics[scale=0.438,  trim=0.0cm 3.0cm 0.0cm 0cm, clip, angle=90]{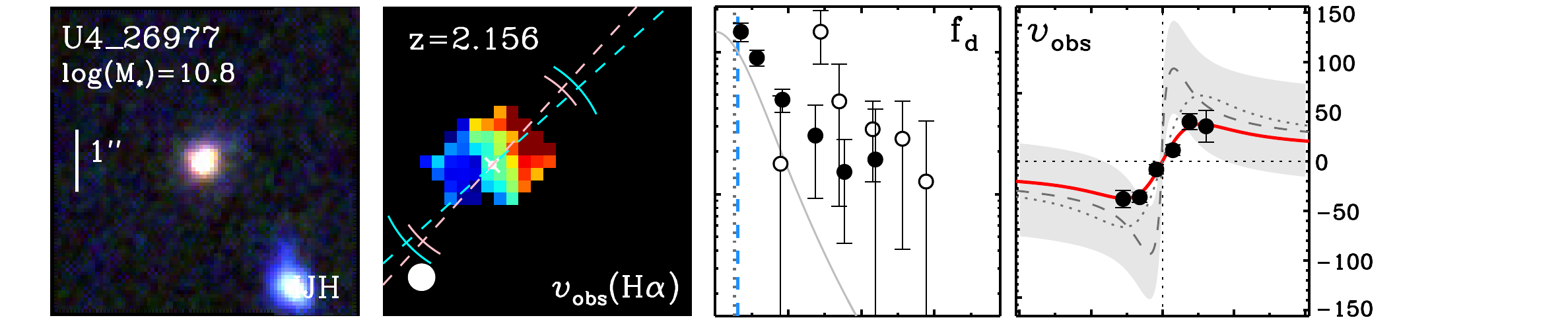}}\\
\vspace*{-1.5em}
\hspace*{-2em}\subfloat{\includegraphics[scale=0.438,  trim=0.0cm 3.0cm 0.0cm 0cm, clip, angle=90]{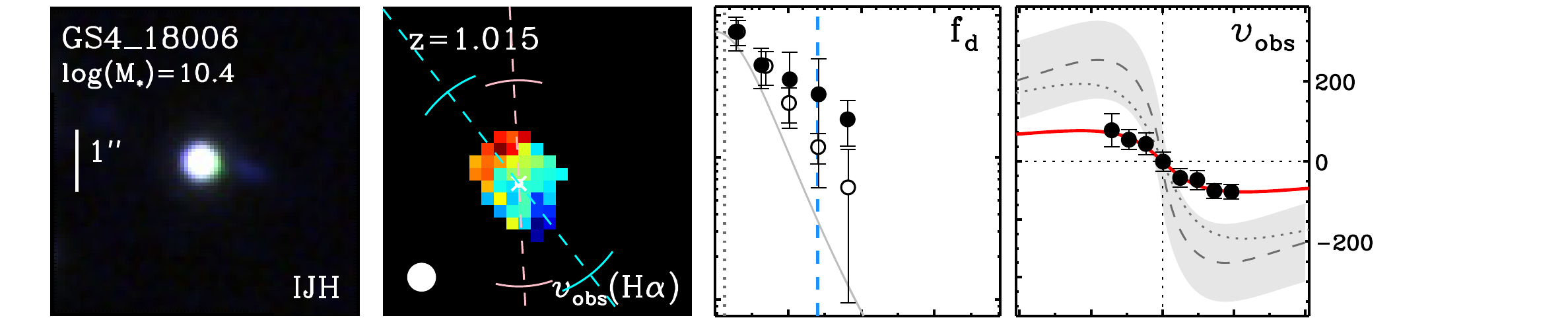}}&\hspace*{-2em}
\subfloat{\includegraphics[scale=0.438,  trim=0.0cm 3.0cm 0.0cm 0cm, clip, angle=90]{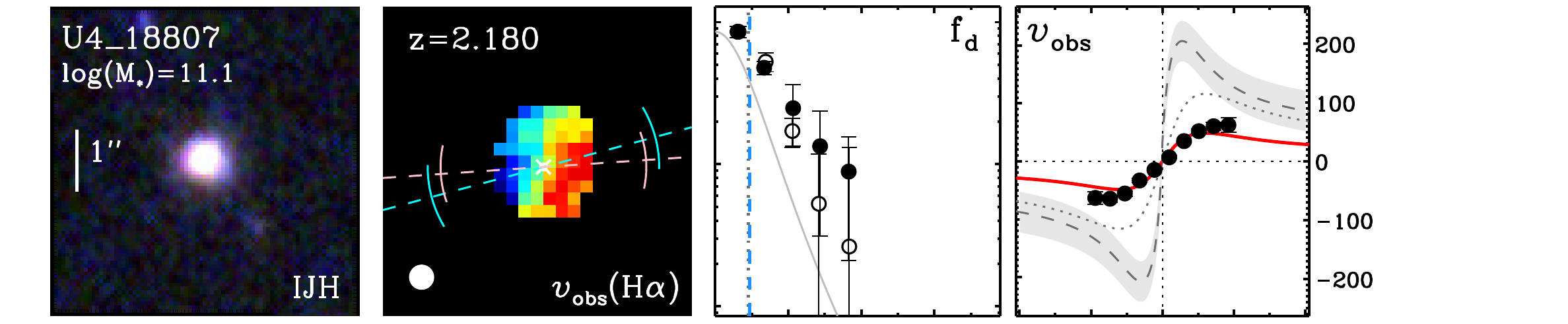}}\\
\vspace*{-1.5em}
\hspace*{-2em}\subfloat{\includegraphics[scale=0.438,  trim=0.0cm 3.0cm 0.0cm 0cm, clip, angle=90]{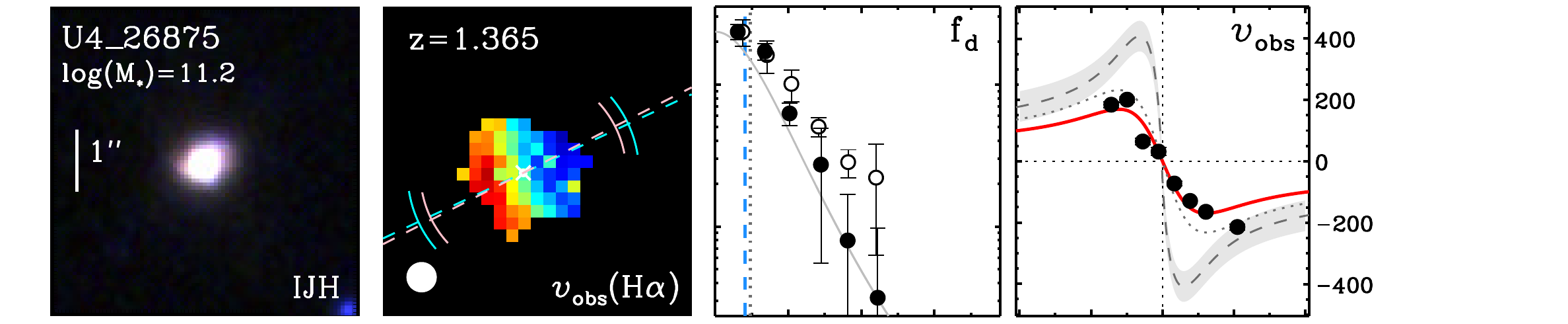}} &\hspace*{-2em}
\subfloat{\includegraphics[scale=0.438,  trim=0.0cm 3.0cm 0.0cm 0cm, clip, angle=90]{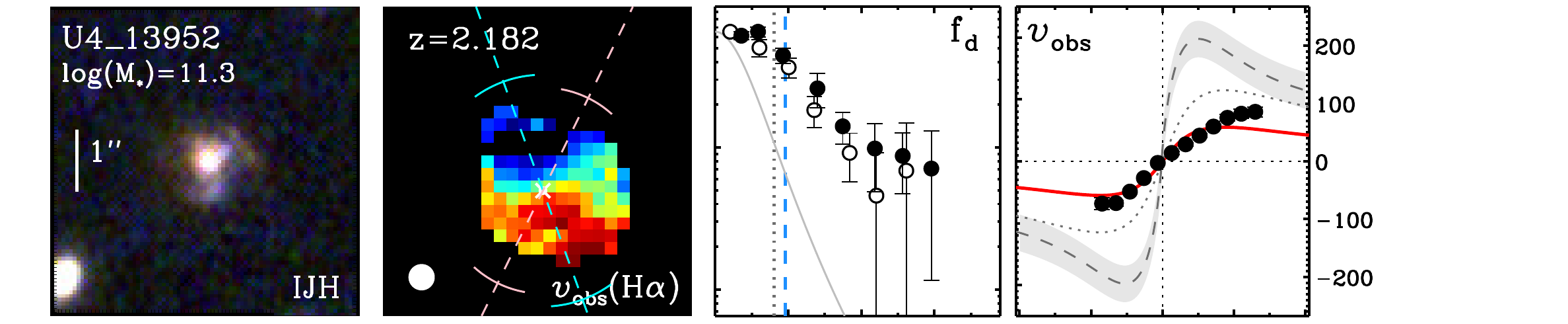}}\\
\vspace*{-1.5em}

\hspace*{-2em}\subfloat{\includegraphics[scale=0.438,  trim=0.0cm 3.0cm 0.0cm 0cm, clip, angle=90]{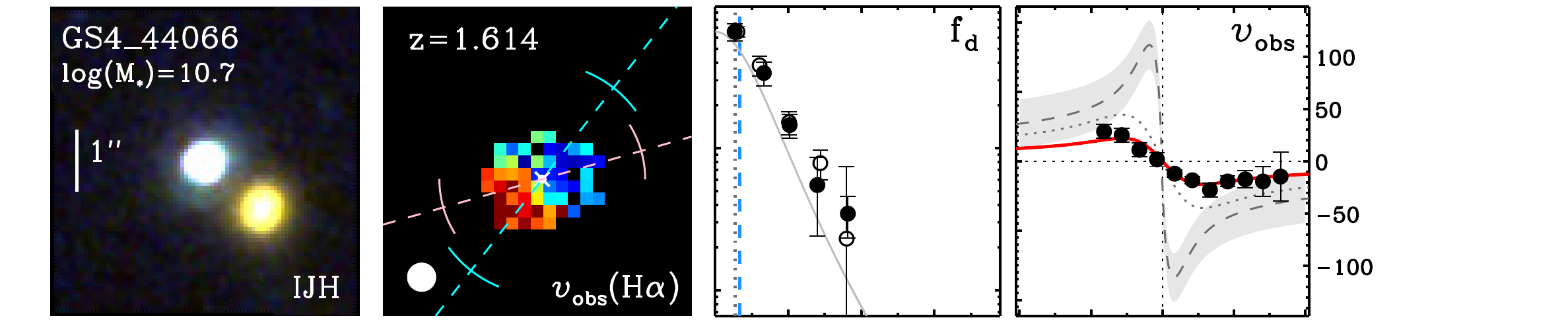}} &\hspace*{-2em}
\subfloat{\includegraphics[scale=0.438,  trim=0.0cm 3.0cm 0.0cm 0cm, clip, angle=90]{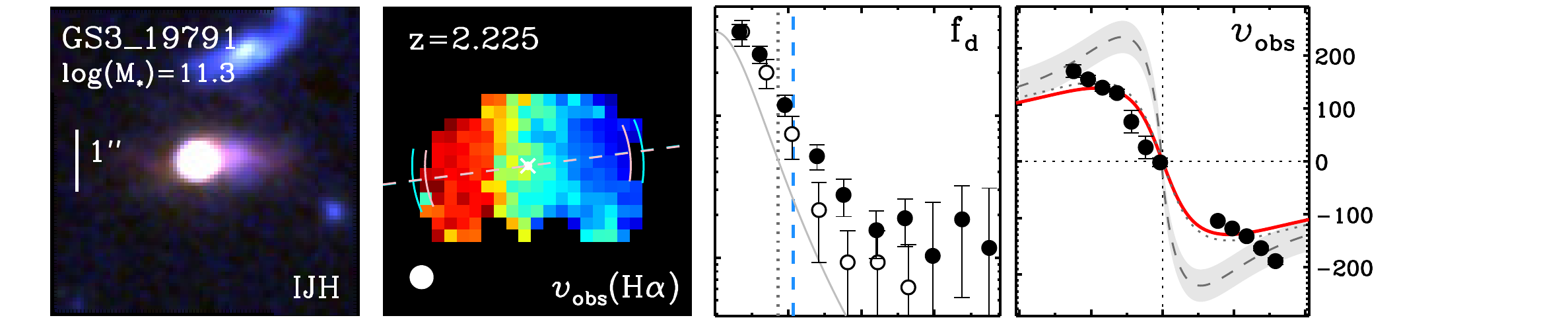}}\\
\vspace*{-1.5em}
\hspace*{-2em}\subfloat{\includegraphics[scale=0.438,  trim=0.0cm 3.0cm 0.0cm 0cm, clip, angle=90]{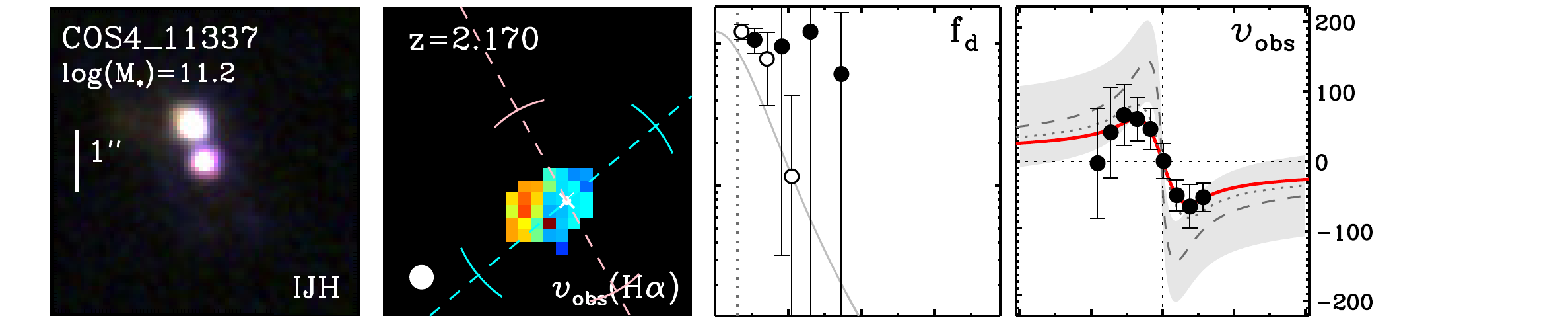}} &\hspace*{-2em}
\subfloat{\includegraphics[scale=0.438,  trim=0.0cm 3.0cm 0.0cm 0cm, clip, angle=90]{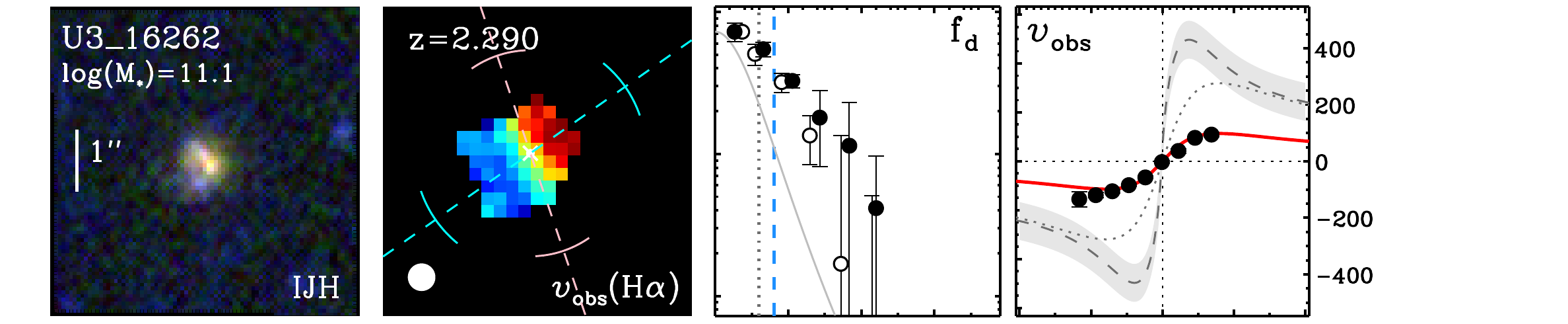}}\\
\vspace*{-1.5em}
\hspace*{-2em}\subfloat{\includegraphics[scale=0.438,  trim=0.0cm 3.0cm 4.1cm 0cm, clip, angle=90]{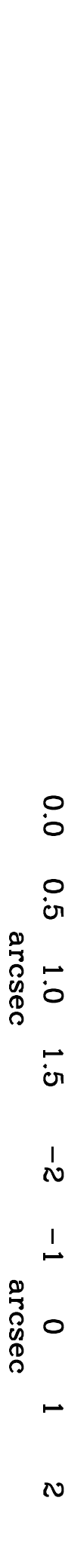}} &\hspace*{-2em}
\subfloat{\includegraphics[scale=0.438,  trim=0.0cm 3.0cm 4.1cm 0cm, clip, angle=90]{axis_kmos_rotcurve_csfg_sm3_min-eps-converted-to.pdf}}
\vspace*{1.5em}

\end{tabular}

\caption[\textwidth]{Extracted 2D and 1D kinematics of all rotationally-dominated compact SFGs in our sample. From left to right: HST $IJH$ color composite image; \kmostd \halpha velocity map shown with FWHM of PSF specific to observations of this galaxy (white circle); normalized \halpha emission (black points) profiles, normalized KMOS continuum (open points) and 1D PSF (gray line) on a logarithmic axis; observed \halpha velocities along major kinematic axis (black points), fit with exponential disk model (red line).  The axis profiles are extracted along the kinematic PA as denoted by the light blue line over plotted on the velocity map. The photometric PA, as determined by F160W HST images, is shown by the pink line. The blue arcs correspond to $\pm18$ degrees, the average misalignment between photometric and kinematic PAs, while the pink arcs correspond to $\pm3\sigma$ error on the photometric PA. In the third panels the half-light radii measured from the $H$-band (dotted gray) and \halpha maps (dashed blue) are shown by vertical lines. In the forth panels the dotted gray velocity curves show the best-ft exponential disk model with the inclination correction applied. The dashed gray velocity curve shows the intrinsic rotation curve. The associated shaded region shows the error on the rotational velocity, $v_\mathrm{rot,corr}$, corrected for both inclination and beam-smearing effects.  }
\label{fig.example}
\end{figure*}

% THESIS: Do cSFGs rotate? have disk-components?
%%%%%%%%
We resolve line emission in 23 of the 35 compact SFGs detected in \kmostdns, spanning the wide redshift range of $0.9<z<3.7$. The fraction of resolved compact SFGs, $\sim66$\%, is marginally lower than the resolved fraction in the full \kmostd sample of 74\%. In Section 4.1 of W15 we outline in detail a set of five criteria used for the \kmostd survey to classify galaxies as `rotation-dominated' and `disk-like'. The criteria are (1) a clear monotonic velocity gradient, (2) the ratio of rotational support to the disk velocity dispersion measured in the outer parts of the galaxy, $v_\mathrm{rot}/\sigma_0$, to be greater than unity, (3) the agreement of the photometric and kinematic axes within 30$^\circ$, (4) the spatial coincidence of the centroids of the velocity map and velocity dispersion map within the errors, and (5) the spatial coincidence of the centroids of the velocity map and the continuum map within the errors. Of the \kmostd compact SFGs that are resolved, {21/23} satisfy the first two criteria to be considered rotation-dominated and {12/23} satisfy the more strict five criteria to be considered disk-like.  These fractions are consistent with the same analysis for the full \kmostd galaxy sample of 83\% and 58\% respectively, as reported in W15.  In Figure~\ref{fig.example} we show the 1D and 2D kinematics of the compact SFGs that are rotation-dominated.

% Figure~\ref{fig.example} shows the observed velocities extracted along the kinematic major axis
\addtocounter{figure}{-1}

\begin{figure*}

\begin{tabular}{ccc}
\vspace*{-1.5em}
\hspace*{-2em}\subfloat{\includegraphics[scale=0.438,  trim=0.0cm 3.0cm 0.0cm 0cm, clip, angle=90]{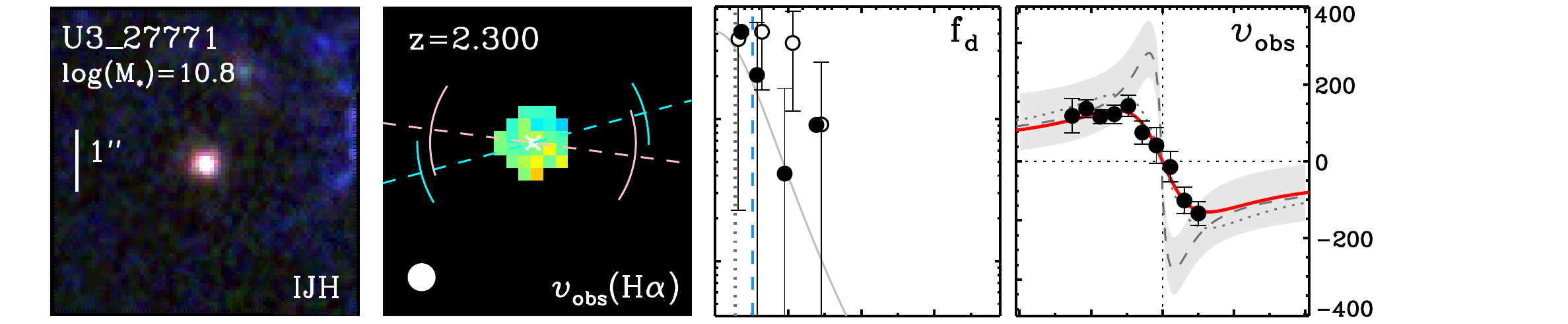}} &\hspace*{-2em}
\subfloat{\includegraphics[scale=0.438,  trim=0.0cm 3.0cm 0.0cm 0cm, clip, angle=90]{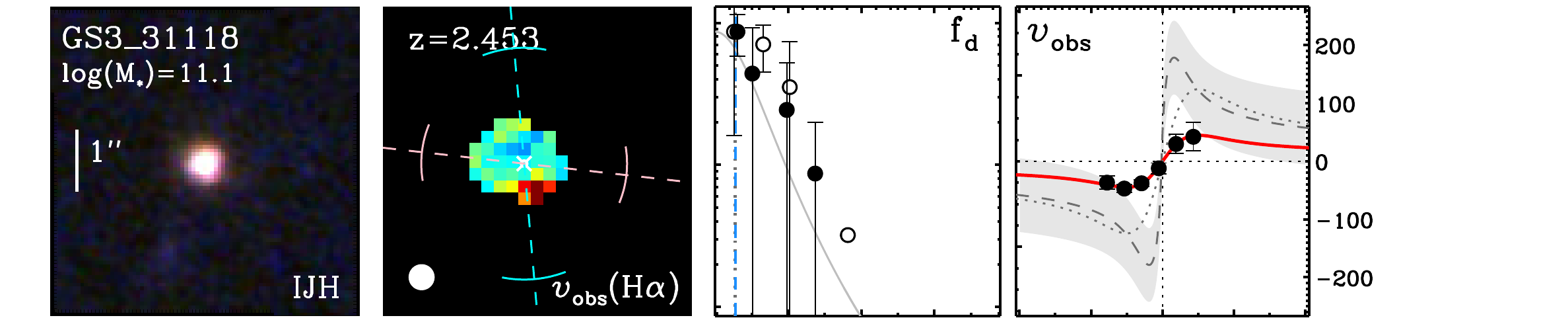}}\\
\vspace*{-1.5em}

%\hspace*{-2em}\subfloat{\includegraphics[scale=0.438,  trim=0.0cm 3.0cm 0.0cm 0cm, clip, angle=90]{minimal/GS3_28008_kmos_rotcurve_csfg_sm3_min.eps}} &\hspace*{-2em}

\hspace*{-2em}\subfloat{\includegraphics[scale=0.438,  trim=0.0cm 3.0cm 0.0cm 0cm, clip, angle=90]{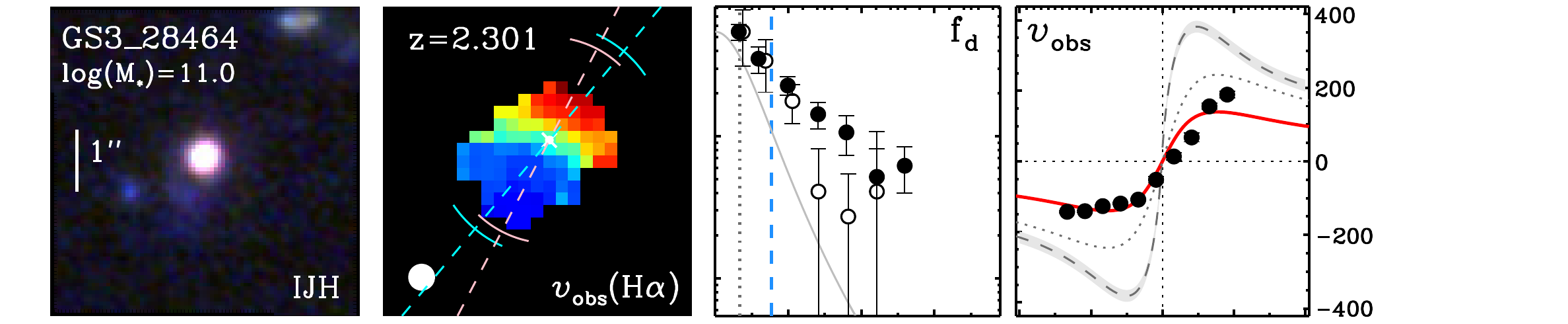}} &\hspace*{-2em}
\subfloat{\includegraphics[scale=0.438,  trim=0.0cm 3.0cm 0.0cm 0cm, clip, angle=90]{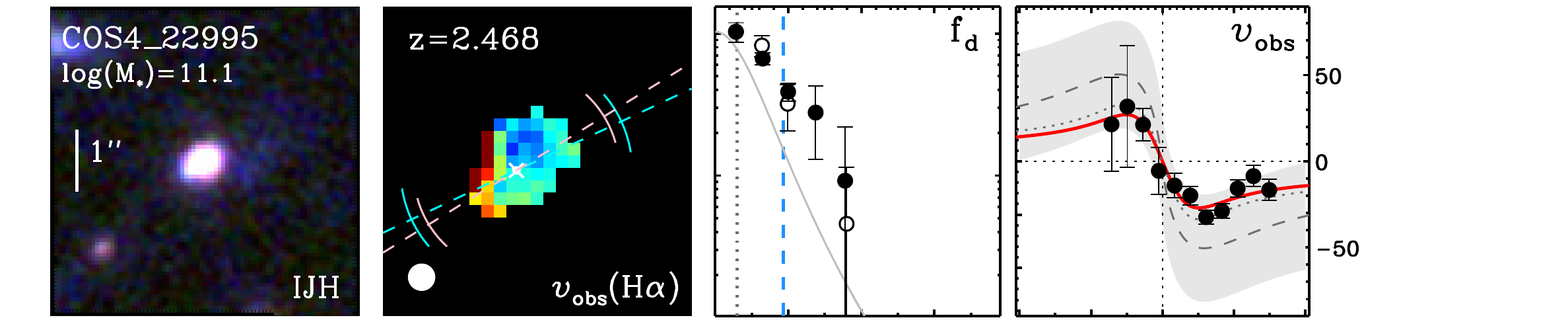}}\\
\vspace*{-1.5em}

\hspace*{-2em}\subfloat{\includegraphics[scale=0.438,  trim=0.0cm 3.0cm 0.0cm 0cm, clip, angle=90]{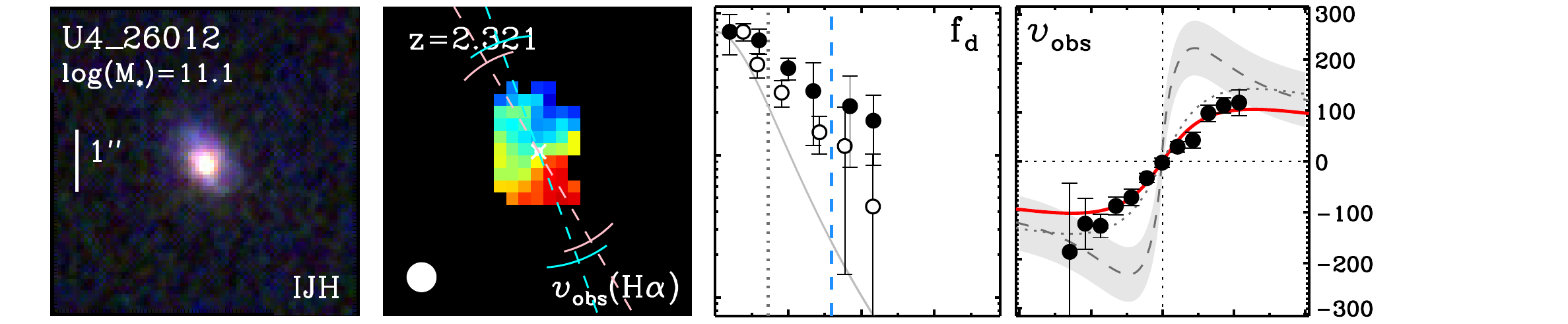}} &\hspace*{-2em}
\subfloat{\includegraphics[scale=0.438,  trim=0.0cm 3.0cm 0.0cm 0cm, clip, angle=90]{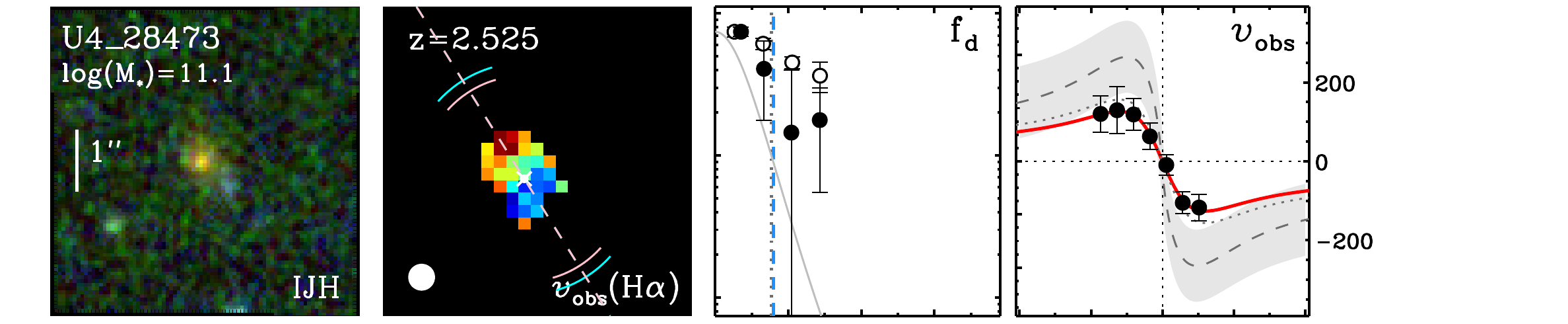}}\\
\vspace*{-1.5em}

\hspace*{-2em}\subfloat{\includegraphics[scale=0.438,  trim=0.0cm 3.0cm 0.0cm 0cm, clip, angle=90]{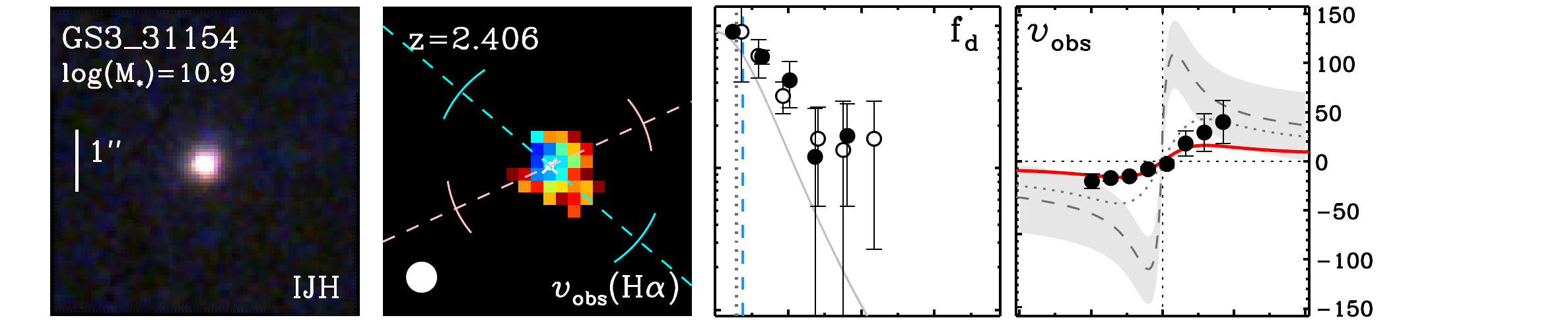}} &\hspace*{-2em}
\subfloat{\includegraphics[scale=0.438,  trim=0.0cm 3.0cm 0.0cm 0cm, clip, angle=90]{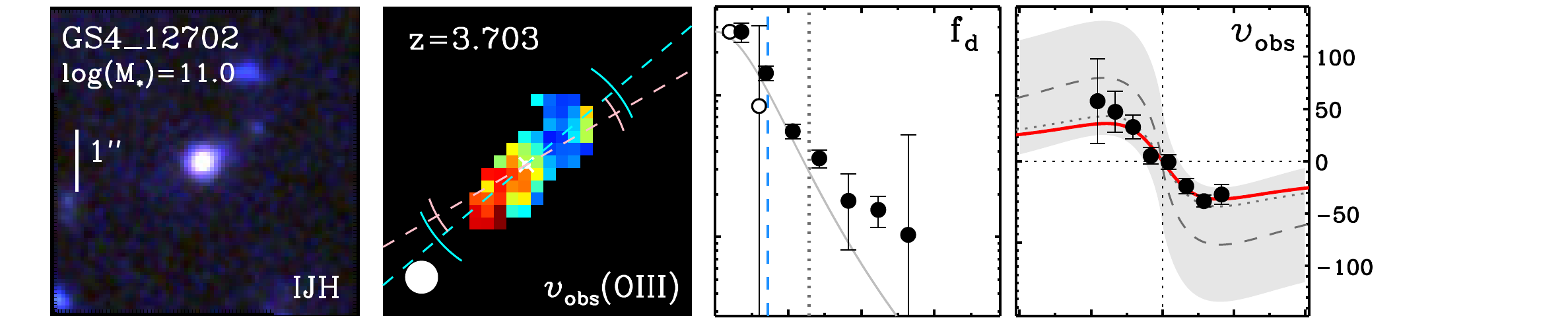}\vspace*{-12cm}}\\
\vspace*{-1.5em}

\hspace*{-2em}\subfloat{\includegraphics[scale=0.438,  trim=0.0cm 3.0cm 0.0cm 0cm, clip, angle=90,valign=T]{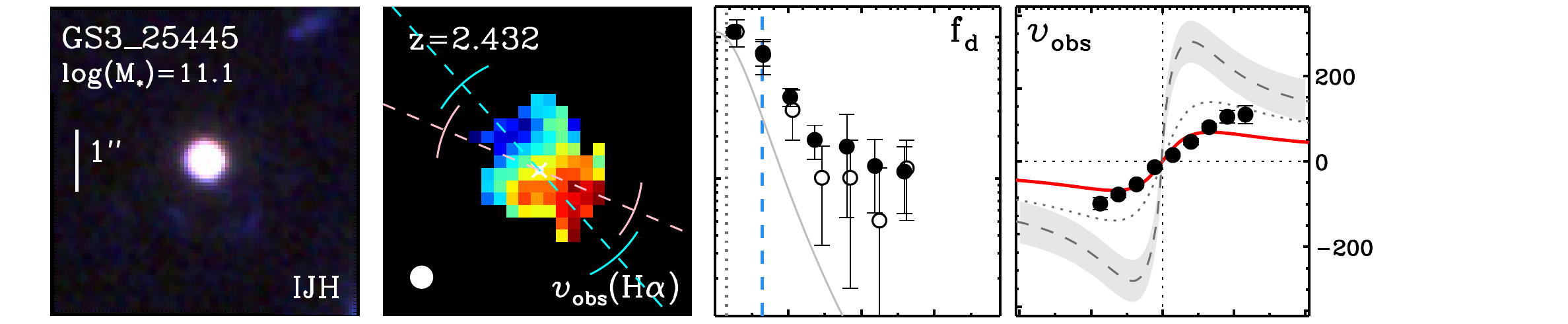}} &\hspace*{-2em}
\subfloat{\includegraphics[scale=0.438,  trim=0.0cm 3.0cm 4.1cm 0cm, clip, angle=90,valign=T]{axis_kmos_rotcurve_csfg_sm3_min-eps-converted-to.pdf}}\\
\vspace*{-1.5em}

\hspace*{-2em}\subfloat{\includegraphics[scale=0.438,  trim=0.0cm 3.0cm 3.8cm 0cm, clip, angle=90]{axis_kmos_rotcurve_csfg_sm3_min-eps-converted-to.pdf}} &\hspace*{-2em}
\vspace*{1.5em}

\end{tabular}

\caption{Kinematic maps and axis profiles for the compact SFGs in \kmostdns, continued.}
\end{figure*}

Rotationally-supported compact SFGs have mean inclination and beam-smearing corrected velocities of {$267$} \kms~and mean {$v_\mathrm{rot,corr}/\sigma_\mathrm{0,corr}=4.6$}, comparable to extended disk-like SFGs from the full \kmostd sample. We discuss these properties in the context of the overall sample in Section~\ref{sec.linewidths}.  Typical uncertainties on $v_\mathrm{rot,corr}$ and $\sigma_\mathrm{0,corr}$ are 25\% and 30\% respectively. The main uncertainties for kinematic measurements of these compact galaxies are discussed further at the end of this section. Disk circular velocities are estimated for the galaxies satisfying all five disk criteria by correcting for the effects of inclination, beam smearing, and additional pressure support from random motions, such that;
\begin{equation}
v_\mathrm{d} = \sqrt{{v_\mathrm{rot,corr}^2}+\alpha\sigma_\mathrm{0,corr}^2}.
\label{eq.vd}
\end{equation}
The functional form of $\alpha$ is dependent on the distribution of gas surface density and gas velocity dispersion \citep{2007ApJ...657..773V,2010ApJ...721..547D,2010ApJ...725.2324B}.  In the case of constant isotropic velocity dispersion \citep{2009ApJ...697..115C} adopted for this paper $\alpha$ is defined as twice the ratio of half-light radius to disk radius, $\alpha=2r_{e}/r_d=3.36$ assuming $v$ is measured at $r=r_e$ \citep{2010ApJ...725.2324B,2016ApJ...826..214B}. We find a range of disk circular velocities, {$v_\mathrm{d}=110-500$ \kms}, for the compact SFGs which follows the underlying \kmostd SFG population. 

Although the kinematic values, $v_\mathrm{rot}$, $v_\mathrm{rot,corr}$, and $v_\mathrm{d}$ are calculated from the maximal velocity difference from the extracted 1D velocity profile, an exponential disk rotation curve is fit for illustrative purposes to the observed 1D velocity profile shown by the red line in Figure~\ref{fig.example}. The disk scale length of the model curve is constrained to be within 2$\sigma$ errors of the \halpha or \OIII~half-light radius. The emission line half-light sizes provide a better prior to obtain a best fit to the observed velocity data in the majority of galaxies. When the radius is left as a free parameter in the rotation curve fit, the resulting radii are more closely matched to the \halpha sizes than the HST $H$-band sizes. Both sizes are shown by vertical lines in the third panels of Figure~\ref{fig.example}.

A further consequence of the small characteristic sizes and morphologies of the compact SFGs is large uncertainties from both the beam smearing corrections, and the $\sin{i}$ corrections to the observed velocity or linewidth. As a result of these corrections the errors on the disk circular velocity for the compact SFGs are large. Even when rotation is resolved the measured $v_\mathrm{obs}$ is a lower limit. Corrections for inclination, ($\sin{i}$)$^{-1}$, are $>2$ for nine compact SFGs due to the high axis ratios characteristic of compact SFGs. For galaxies with axis ratios very close to unity the true inclination correction is highly uncertain. This uncertainty is propagated to $v_\mathrm{d}$, however the $\sim4-10$\% errors on the axis ratio may be underestimated for the compact sources presented in this work. In Figure~\ref{fig.example} the dotted line shows the inclination correction applied to the best-fit model curve to the observed data. 

A beam smearing correction is applied to the observed velocity, $v_\mathrm{obs}$, in addition to the inclination correction. The beam smearing corrections depend on the ratio of the $H$-band effective radius to the KMOS PSF. As a result, galaxies with compact $H$-band sizes have large correction factors. In Figure~\ref{fig.example} the intrinsic non-beam-smeared rotation curve assuming the exponential disk radius is equal to $r_\mathrm{e}[F160W]/1.68$ is shown by the dashed line. The average velocity beam smearing correction factor is 1.5, with a range from $1.2-1.9$. However, as discussed, inferred intrinsic \halpha sizes can be $1-4\times$ greater than the $H$-band sizes. In these cases the beam-smearing corrections may be over-estimated when using the $H$-band size.  Beam smearing corrections are calculated using the \halpha size for 6/23 resolved galaxies in the sample corresponding to the galaxies with \halpha sizes 2-sigma larger than their $H$-band sizes. For these 5 galaxies, the intrinsic non-beam-smeared rotation curve assumes that the exponential disk radius is equal to $r_{\mathrm{H}\alpha}/1.68$. The grey band surrounding the dashed line reflects the errors on the observed velocity, inclination correction, and beam-smearing corrections. 

Errors on the beam smearing corrections are estimated from Monte Carlo simulations of the galaxy parameters that enter into the beam smearing calculations. For the velocity beam smearing correction only the half-light radius is varied. The resulting 34 and 68 percentile errors on the velocity beam smearing correction are small, typically a few percent. The dominant correction to the velocity is galaxy dependent as can be seen from the variety of rotation curves in Figure~\ref{fig.example}. Beam smearing corrections to $\sigma_0$ are dependent on $M_*$, $i$, and $r_\mathrm{e}$ as detailed in Appendix A.2.4 of \cite{2016ApJ...826..214B}. Multiplicative corrections range from $0.2-0.9$.  The 34 and 68 percentile errors on the dispersion beam smearing correction are larger, typically 40\%. 

A higher fraction of the compact galaxies may be rotating but observations are limited by beam size and low surface brightness (despite pushing to low flux levels for many cases, $\sim 4\times10^{-18}$ ergs s$^{-1}$ cm$^{-2}$ arcsec$^{-2}$ using integration times $>8$ hrs).  {Deep AO-assisted data (updating previously seeing-limited observations) of small SFGs has revealed that previously classified dispersion-dominated SFGs ($v_\mathrm{rot}/\sigma_0<1$) are actually rotationally supported \citep{2013ApJ...767..104N}}. Initial observations of compact SFGs yielded large linewidths, $\sigma_\mathrm{tot}\sim200-300$ \kms~\citep{2014ApJ...795..145B,2014Natur.513..394N}, in some cases interpreted under the assumption that compact SFGs have negligible rotation \citep{2014ApJ...795..145B}. Recent results, also from the 3D-HST sample \citep{2015ApJ...813...23V}, find a wide range of linewidths and argue that rotation likely \textit{does} provide the majority of dynamical support for these galaxies. We directly measure with integral field spectroscopy a similar but wider range of rotational velocities than those in \cite{2015ApJ...813...23V}.  Compact SFGs have axis ratios close to unity resulting in poorly constrained photometric axes. An advantage of resolved spectroscopy for these galaxies is that the kinematic axis can be measured independent of the photometric axis. Approximately half, {11/21}, of the rotation-dominated compact SFGs have a kinematic misalignment from the photometric axis of $>30$ degrees as seen by the blue and pink lines in the second panels of Figure~\ref{fig.example}. 

 \begin{figure*}[thbp]
\begin{center}
\includegraphics[scale=0.78, trim=0cm 0cm 0cm 0.7cm, angle=90]{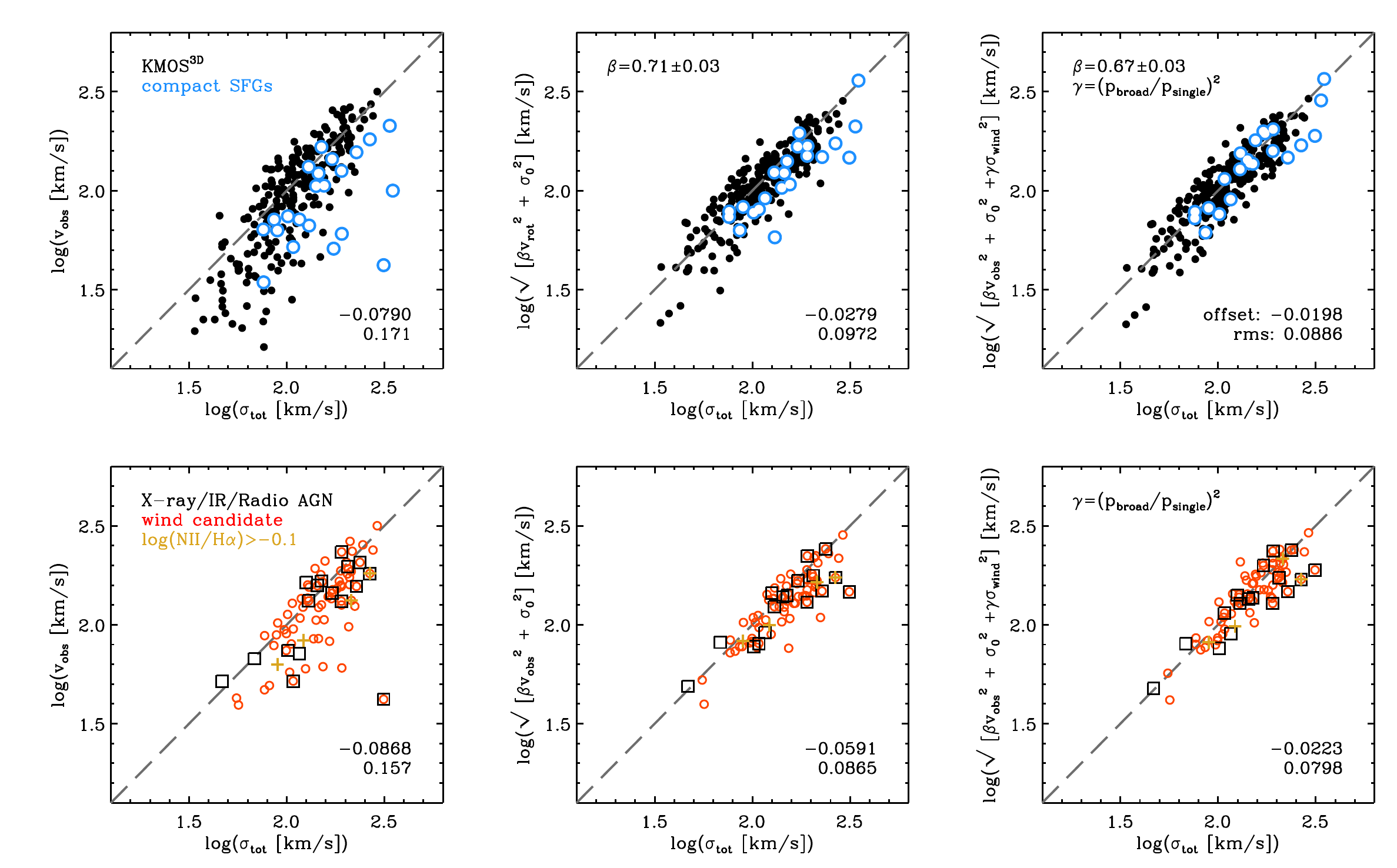}
\caption{Combinations of resolved kinematic parameters compared to the observed galaxy-integrated linewidths using equation~\ref{eq.linewidth}. The top panels show the full sample of \kmostd galaxies that have rotation with obvious pairs and mergers removed (black symbols). Compact SFGs are identified in blue. The bottom panels show the same comparisons for only the candidate AGN host galaxies as selected by X-ray/IR/radio techniques (black squares; see \citealt{2014ApJ...796....7G}, F\"orster Schreiber et al.~\textit{in prep}), the presence of a secondary broad \halpha component from the central regions (red circles), or with $\log$(\NII/\halphans)$>-0.1$ (orange crosses). The numbers on the bottom right corner of each panel give the mean offset and scatter in dex respectively. }
\label{fig.linewidth}
\end{center}
\end{figure*}

\subsection{The dynamics of integrated linewidths}
\label{sec.linewidths}
As discussed in Section~\ref{sec.unresolved} the measured integrated emission linewidth represents the total dynamics of the system including both rotation and dispersion as well as non-gravitational effects such as galactic-scale winds. With the resolved information from KMOS we decompose the relative contributions from turbulence, rotation, and large-scale winds to the measured \stot~of compact and extended SFGs with direct measurements. 

In theory the observed linewidth should be a linear combination of these components, such that
\begin{equation}
\sigma_\mathrm{tot} \approx  \sqrt{\beta v_\mathrm{obs}^2 +\sigma_\mathrm{ISM}^2 +\gamma\sigma_\mathrm{wind}^2},
\label{eq.linewidth}
\end{equation}
 (e.g. \citealt{1985ApJS...58...67T,2006ApJ...653.1027W,2015ApJ...813...23V}). In equation~\ref{eq.linewidth}, $\beta$ is a constant that we calibrate below, $\sigma_\mathrm{ISM}$ is the dispersion of the interstellar medium of the galaxy here approximated by $\sigma_0$, and $\sigma_\mathrm{wind}$ is the dispersion of broad underlying emission from a large scale wind scaled by factor $\gamma$. The functional form of $\gamma$ is unknown with likely dependence on galaxy inclination and wind opening angle. In the absence of knowledge of the detailed structure of the winds we parameterize $\gamma$ empirically as the square of the ratio of the peak of the broad component in a two-Gaussian fit to the peak of a single Gaussian fit. This is based on the assumption that the closer the ratio is to unity the larger the effect of the wind on the total linewidth.  Here, in contrast to equation~\ref{eq.vd}, the ISM turbulence, measured as $\sigma_0$, is added in quadrature with no scale factor, reflecting its contribution to line-broadening, and no beam smearing correction is applied. Thus, in the case of a purely face-on disk with no winds the integrated linewidth is equivalent to the isotropic disk velocity dispersion, as both observables $v_\mathrm{obs}$ and \stot~are affected by inclination. 

In Figure~\ref{fig.linewidth} we test the validity of equation~\ref{eq.linewidth} for $z\approx1-3$ \kmostd galaxies showing rotation (disk criterion 1). In the top left panel of Figure~\ref{fig.linewidth} we show the measured velocity difference, $v_\mathrm{obs}$, as a function of the observed linewidth, $\sigma_\mathrm{tot}$, for compact SFGs as they relate to the full population of \kmostd galaxies. While the full sample shows a general agreement at  high observed velocities and linewidths, there is significant scatter with the tendency for the velocity shear to represent only a fraction of the linewidth for both the compact SFGs and full population. This is unsurprising as local gas velocity dispersions are known to be high at these redshifts (e.g. \citealt{2006Natur.442..786G,2006ApJ...645.1062F,2007ApJ...669..929L,2009ApJ...706.1364F, 2009ApJ...697..115C,2012ApJ...758..106K,2013ApJ...767..104N,2016MNRAS.457.1888S}; W15).  Adding the turbulent motions in quadrature to the observed velocity difference (top middle panel) reduces the rms comparison to the observed linewidth for the full \kmostd sample of rotating galaxies from 0.171 dex to 0.097 dex with a best-fit scale factor of $\beta=0.71\pm0.03$. Accounting for turbulence is fractionally most important in systems with low $v_\mathrm{obs}$ of 100 \kms~or less including near face-on objects and galaxies with $v/\sigma\sim1-2$. % \textbf{\color{BrickRed}While we are able to confirm with our observations that compact SFGs can be rotation-dominated systems, the top panels of Figure~\ref{fig.linewidth} shows intrinsic disk dispersion in compact SFGs is non-negligible with it accounting for 50\% of the linewidth on average. making a direct inference of  rotation velocity from a linewidth highly uncertain.}

The right-most panels of Figure~\ref{fig.linewidth} show the comparison with observed integrated linewidth when the contribution of an additional non-gravitational component is considered. For galaxies in our sample with strong broad components we make a two-Gaussian fit to the rotation-corrected integrated spectrum to determine the contribution from nuclear- or SF-driven winds to the integrated linewidth (\citealt{2014ApJ...787...38F, 2014ApJ...796....7G}, F\"orster Schreiber et al.~\textit{in prep}).  For the strongest cases an additional broad Gaussian component, $\sigma_\mathrm{wind}$, produces a significant improvement to the overall model spectral fit. Using a conservative cut, we measure $\sigma_\mathrm{wind}$ in {68} galaxies, a quarter of our sample, that also have $\sigma_0$ and $v_\mathrm{obs}$ measurements. We see a further reduction of scatter in Figure~\ref{fig.linewidth} when taking the wind component into account, with the ratio of the peak emission of the broad and single components $\gamma=(p_\mathrm{broad}/p_\mathrm{single})^2$ and a best-fit scale factor of $\beta=0.67\pm0.03$. Although the overall effect is small for the full resolved \kmostd sample, the importance for the compact SFGs can be seen from the offset in the top middle and right panels of Figure~\ref{fig.linewidth}.

The bottom panels of Figure~\ref{fig.linewidth} show the same linewidth comparisons for only AGN candidates using all available AGN indicators. The wind term of equation~\ref{eq.linewidth} with scale factor $\gamma$ is particularly important for the subsample of galaxies that show evidence of hosting nuclear-driven outflows, many of which are classified as compact SFGs. For the galaxies possibly hosting AGN we see comparable trends as for the full rotation-dominated sample with a reduction in scatter from 0.087 dex to 0.080 dex when including the contributions from winds indicating that the presence of an AGN could influence the derived 1D kinematics as seen from the spectra shown in Figure~\ref{fig.intspec1} of COS4\_11363 for example.

%solving for beta
The velocity scale factor, $\beta$, in equation~\ref{eq.linewidth} reflects the projection of the rotational velocity along the line-of-sight with literature values ranging from $0.5-0.75$ (e.g. \citealt{1997MNRAS.285..779R, 2006ApJ...646..107E,2006ApJ...653.1027W,2015MNRAS.450.2327Z}). We use our observables to solve for the value of $\beta=0.67\pm0.03$, which brings the left and right sides of equation~\ref{eq.linewidth} into agreement for the \kmostd data as shown in the right most panels of Figure~\ref{fig.linewidth}. If the wind component is not considered (middle panels) then $\beta$ is slightly lower but consistent within the uncertainties, 0.71$\pm0.03$. This value is higher than typically used by previous studies, however $\beta$ may encompass a mix of dependencies on aperture size, line-profile asymmetries, beam-smearing, centrally weighted light-profiles, and the magnitude of random motions \citep{1997MNRAS.285..779R} and thus may be sample or data specific.

%relation to s0.5
The exact form of equation \ref{eq.vd} and equation \ref{eq.linewidth} and the constant $\alpha$ in equation~\ref{eq.vd} is dependent on the radial distribution of density and velocity dispersion assumed. For a \textit{spherically} symmetric system with isotropic velocity dispersion $v=\sqrt{2}\sigma$ or $v=\sqrt{3}\sigma$ depending on the exact model used \citep{2008gady.book.....B}. This formalism motivates the $S_{0.5}$ dynamical parameter popularized for use in Tully-Fisher analyses (e.g. \citealt{2006ApJ...653.1027W,2007ApJ...660L..35K,2014ApJ...795L..37C}). Setting $\beta=0.5$ here produces a similar reduction in scatter (rms=0.0995) but leads to a larger offset with 78\% of values having $\sigma_\mathrm{tot}>S_{0.5}$ (offset=-0.0756). 

%caveat
A caveat is that in small or faint galaxies in which the KMOS observations do not probe to the outer regions of the disk, the $v_\mathrm{obs}$ values may be biased low and the $\sigma_0$ values may be biased high. However, the dual measurement effects act together minimising their effect \citep{2010ApJ...710..279C}.

\subsection{Integrated line ratios}
\begin{figure}[t]
\begin{center}
\includegraphics[scale=0.77, angle=90, trim=0cm 0.4cm 0.6cm 0.5cm, clip]{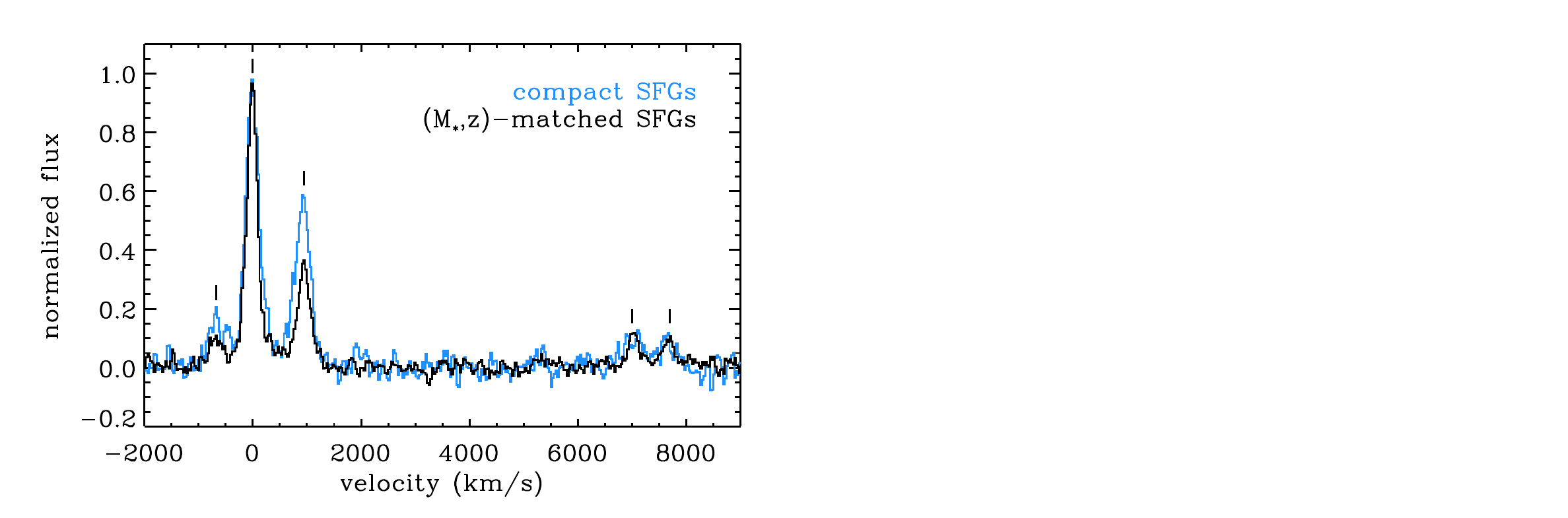}
\caption{Stacked KMOS spectra for 23 compact SFGs (blue) and a stellar mass and redshift matched comparison sample of 46 galaxies from \kmostd (black). The velocity axis is centered around \halpha. The positions of \halphans, \NII$\lambda\lambda6548,6584$, and \SII$\lambda\lambda6717,6731$ are identified with vertical lines. }
\label{fig.stack}
\end{center}
\end{figure}

%describe stacking
We measure a wide range of $\log$(\NII/\halphans) ratios, $-0.5-0.2$, for compact SFGs as shown in Figure~\ref{fig.intspec1}. We combine the spectra in a median stack of \halpha normalized spectra to determine the average emission line properties of compact SFGs. By stacking we are able to detect the weaker lines \SII$\lambda\lambda6717,6731$. Broad line AGNs are removed prior to stacking. For comparison a mass and redshift matched stack is also produced drawing from the full \kmostd sample. Both stacked spectra are shown in Figure~\ref{fig.stack}. The residuals from a single component Gaussian fits to the \halphans-\NII-\SII~lines reveal excess emission especially near the \halphans-\NII~complex. This emission is likely from stellar or nuclear driven winds as discussed above. For simplicity we adopt a single component fit to the emission lines. However, we note that the narrow component for a two component fit is consistent within $2\sigma$ errors with the single component fit.

%describe results
The compact SFGs (blue spectrum in Figure~\ref{fig.stack}) show a higher \NII/\halpha ratio, $0.58\pm0.2$, than the mass and redshift matched comparison sample (black spectrum), $0.38\pm0.1$ while the \SII/\halpha ratios are consistent within the errors, $0.21\pm0.2$ and $0.22\pm0.1$ respectively. The high \NII/\halpha ratios are expected given the high AGN fraction in massive galaxies and particularly the compact SFGs as discussed in the previous section. The low \SII/\halpha ratios in both stacks may be the result of a higher ionization parameter \citep{2008MNRAS.385..769B,2015ApJ...812L..20K,2017ApJ...835...88K} and are inconsistent with pure shock or LINER driven emission (from local calibrations; \citealt{2011ApJ...734...87R}).

%describe implications and rehash in discussion/conclusions
\subsection{Dynamical mass}

We estimate the dynamical mass, \mdyn, assuming a thick (1:4) exponential disk for the \kmostd rotation-dominated galaxies (disk criterion 1 and 2) from the rotational velocities using:
\begin{equation}
M^d_\mathrm{dyn}(r<r_e) =\frac{v_\mathrm{d}^2r_e}{\alpha_0 G}\mathrm,
\label{eq.mdynV}
\end{equation}
where the factor $\alpha_0$ is dependent on the mass distribution of the system \citep{2008gady.book.....B}. We assume $\alpha=1.09$ for a pressure supported thick disk. We use the half light sizes, $r_e$, measured from observed F160W $H$-band light from the CANDELS imaging using single \sersic fits \citep{2014ApJ...788...11L}. For a full analysis of the baryon fractions of \kmostd galaxies and a 2D modeling approach see \cite{2016ApJ...831..149W}. The dynamical mass estimates would marginally increase if \halpha half-light sizes were used for the full sample as $r_{\mathrm{H}\alpha}/r_e$ is close to unity for the majority of star-forming galaxies at this epoch (\citealt{2014Natur.513..394N}, Wilman et al.~\textit{in prep}).%See also Bertrin et al. 2002, Cappelari et al. 2006. 

For galaxies with unresolved kinematics we estimate the disk circular velocity from the observed linewidth, under the assumption that their dynamics are dominated by rotation \citep{2008gady.book.....B}. Because disk velocity dispersions are high at these epochs proving non-negligible pressure support, equation~\ref{eq.linewidth} alone is not sufficient to derive the velocity. Thus to derive the disk circular velocity from the integrated linewidth the corrections take the composite form of;
\begin{equation}
v_\mathrm{d} = \sqrt{ \frac{(\sigma_\mathrm{tot}^2 - \sigma_\mathrm{0}^2 -\gamma\sigma_\mathrm{wind}^2)}{\beta\sin^2{i}} +3.36\sigma_0^2}.
\label{eq.vdlinewidth}
\end{equation}

\begin{figure}[thbp]
\begin{center}
\includegraphics[scale=0.8, angle=90, trim=0cm 0cm 11.2cm 13cm, clip]{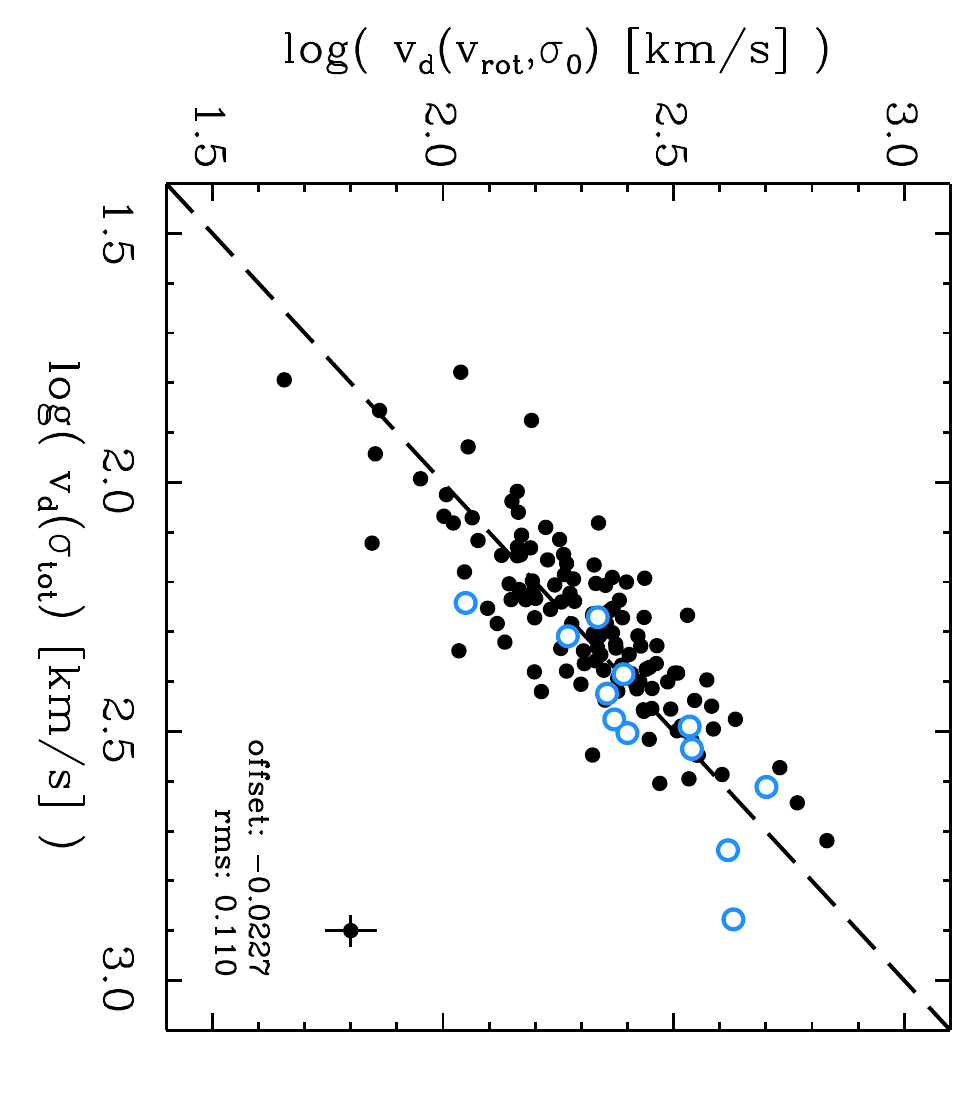}
\caption{  Disk circular velocity derived from the resolved kinematic parameters, $v_\mathrm{rot}$ and $\sigma_0$, versus the disk circular velocity derived from the integrated linewidth using equation~\ref{eq.vdlinewidth}. Symbols are the same as in Figure~\ref{fig.linewidth}. A representative error bar is shown in the bottom right corner. }
\label{fig.vdstot}
\end{center}
\end{figure}

However, equation~\ref{eq.vdlinewidth} requires knowledge of $\sigma_0$ and $\sigma_\mathrm{wind}$, which are not available in the unresolved case. We estimate $\sigma_0$ based solely on redshift such that $\sigma_0\sim25$ \kms~at $z<1.2$, $\sigma_0\sim35$ \kms~at $1.2<z<1.8$, and $\sigma_0\sim50$ \kms~at $z>1.8$ (W15) and assume that $\sigma_\mathrm{wind}=0$.  We test this method on resolved rotation-dominated galaxies, shown in Figure~\ref{fig.vdstot}, finding reasonable agreement between $v_\mathrm{d}$ estimated from resolved parameters in equation~\ref{eq.vd} and $v_\mathrm{d}$ estimated from equation~\ref{eq.vdlinewidth}, $v_\mathrm{d} (v_\mathrm{rot},\sigma_0) / v_\mathrm{d} (\sigma_\mathrm{tot})=1.15$ with 0.11 dex scatter. A large portion of the scatter is likely due to the range in measured $\sigma_0$ values at each epoch (W15). 

We also test the simpler case in which we assume that the turbulence and wind contributions are negligible and only apply an inclination correction to \stot. The comparison between inclination corrected \stot~and $v_\mathrm{d}$ shows on average that {$v_\mathrm{d}\sim1.31(\sigma_\mathrm{tot}/\sin{i})$ }for the full sample. Although less physically motivated, the rms scatter for this method is comparable to using the method described above.

\begin{figure}[thbp]
\begin{center}
\includegraphics[scale=0.8, angle=90, trim=10.5cm 12cm 0.8cm 1cm, clip]{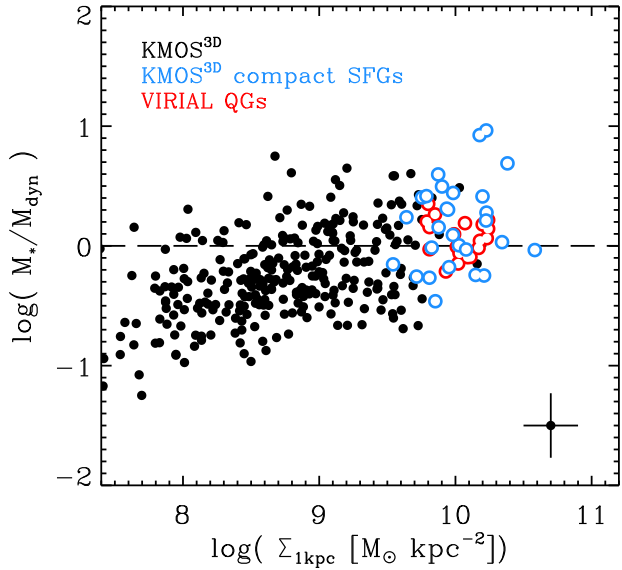}
\caption{Stellar to dynamical mass fraction as a function of inner-kpc density for extended SFGs, compact SFGs, and quiescent galaxies (QGs). Dynamical masses are estimated from equation~\ref{eq.mdynV} for SFGs and from JAM modeling for quiescent galaxies from the VIRIAL survey (\citealt{2015ApJ...804L...4M}; \textit{in prep}). A representative error 1-sigma bar is shown in the bottom right corner. }
\label{fig.mdyn}
\end{center}
\end{figure}

In Figure~\ref{fig.mdyn} we investigate the relationship between stellar and dynamical mass as a function of inner-kpc density recovering a weak trend such that the most centrally dense objects, including compact SFGs and quiescent galaxies, have stellar mass fractions closest to unity. We compare our results to the VIRIAL survey of quiescent galaxies. In contrast, their dynamical masses are computed from a combination of stellar kinematics and photometry with JAM modeling \citep{2008MNRAS.390...71C}. With this comparison sample, we find that compact SFGs are consistent with quiescent galaxies in both central density and stellar baryon fraction with $M_*/M_\mathrm{dyn}$ scattered around unity. This implies that there is little room for a significant additional mass component from molecular gas, atomic gas, or dark matter within the regions of the galaxies probed by our measurements. The high stellar to dynamical mass ratios and high stellar densities suggest a short time scale for the onset of quenching. Indeed, recent CO and [CI] ALMA observations of compact SFGs GS3\_19791 and COS4\_22995, also in our sample, reveal a molecular gas fraction between 4-14\% as well as short depletion times \citep{2016ApJ...832...19S,2017A&A...602A..11P}. These gas fractions are consistent with gas fractions implied by the ratio of the dynamical mass estimates and stellar masses. They are significantly lower than the average gas fractions of `typical' $z\sim2$ extended SFGs of $\gtrsim$40\% \citep{2010Natur.463..781T,2013ApJ...768...74T,2017arXiv170201140T,2010ApJ...713..686D,2015ApJ...800...20G}. 

As discussed in \cite{2016ApJ...831..149W} some galaxies scatter to unphysical baryon fractions. We investigate this further for the specific case of compact SFGs looking for trends with galaxy properties.  Inclination corrections are particularly uncertain for compact SFGs as the typical axis ratios are close to unity. They are difficult to measure due to both the circular appearance of the galaxies as well as their small size in comparison to the HST PSF. The compact SFG with the highest stellar baryon fraction has axis ratio $q\sim0.6$. If the axis ratio was increased to $q=1$, comparable with other compact SFGs of the sample, then $\log{(M_*/M_\mathrm{dyn})}$ would be reduced to $-0.3$, in line with the scatter of the rest of the population. 

%%%%%%%%%%%%%%%%%%%%%%%%%%%%%%%%%%%%%%%%%%%%%%%%
\section{Discussion}
\label{sec.dis}
%%%%%%%%%%%%%%%%%%%%%%%%%%%%%%%%%%%%%%%%%%%%%%%%
%{\color{BrickRed}MAKE SURE THIS SECTION STAYS FOCUSED ON THE KINEMATICS***}
Compact SFGs have garnered a lot of attention in the last five years as a possible transitional population between the massive end of the star-forming main sequence and the quiescent galaxy population (e.g. \citealt{2013ApJ...765..104B,2014ApJ...791...52B,2014ApJ...795..145B,2014ApJ...780....1W,2014MNRAS.438.1870D,2014ApJ...780...77T,2015MNRAS.451.2933B}). Two dominant theories have emerged as to how compact SFGs are formed; either through gas dissipation and central starburst \citep{2014MNRAS.438.1870D,2017ApJ...834..135T} or from already small less-massive SFGs \citep{2015ApJ...813...23V,2016ApJ...833....1L}. In contrast, compact SFGs have universally been linked as the immediate progenitors of compact quiescent galaxies at $z\sim1-2$ and thus the likely progenitors of a fraction of S0 or elliptical galaxies in the local universe. We explore our resolved \halpha results in the context of these scenarios.

%%%%%%%%%%%%%%
\subsection{Forming compact SFGs}
%%%%%%%%%%%%%%
In the `compaction' formation scenario central starbursts within extended galaxies rapidly build-up a central mass concentration creating the possible precursors of compact SFGs. \cite{2017ApJ...841L..25T} present two such galaxies at $z=2.5$ with dust obscured cores and compact molecular gas sizes ($\sim1.3$ kpc). The starburst cores have sizes, $v_\mathrm{rot}/\sigma$, and $v_\mathrm{rot}$ comparable to our \halpha results of compact SFGs. However it follows that a more extended stellar disk remains that may be obscured or out-shined by the central core. The measured \halpha profiles of \kmostd compact SFGs with exponential disks of $r_{\mathrm{H}\alpha}\sim2.5\pm0.2$ kpc and high stellar baryon fractions suggest that we are unlikely missing substantially larger extincted star-forming disks.

Alternatively, the \halpha and continuum sizes may more simply suggest that compact SFGs are amongst the oldest galaxies in our full SFG population such that their current small observed sizes are reflective of the average population at the epoch when they assembled the bulk of their stellar mass (e.g. \citealt{2015ApJ...813...23V,2016ApJ...833....1L}). In this scenario, dissipative processes may still have been responsible for creating the high central density but have done so on a different time scale than the central starburst scenario. 

%%%%%%%%%%%%%%
\subsection{Compact SFGs as progenitors of compact quiescent galaxies}
%%%%%%%%%%%%%%

The high stellar to dynamical mass ratios measured for the \kmostd compact SFGs imply that the onset of quenching of the remaining star-formation should be fast as there is little room for molecular gas reservoirs.
Further support for short timescales comes from the morphology of compact SFGs (e.g  bulge-to-total ratios: B/T $\sim0.5$ and \sersic indices: $n\sim3$). Galactic structure, both locally and at $z\sim1-2$, have been linked to high passive fractions for massive galaxies \citep{2003MNRAS.341...54K,2006MNRAS.368..414D,2011ApJ...742...96W,2012ApJ...753..167B}.  The results presented here and in other \kmostd papers \citep{2016ApJ...826..214B,2016ApJ...831..149W} are consistent with scenarios in which SFGs approach a critical mass or mass density before quenching.

Our kinematic results reveal that $>50\%$ of compact star-forming galaxies have a disk component with significant rotational support. If compact SFGs are the true progenitors of quiescent galaxies then the quenching process at this epoch will either destroy the rotation leaving a pressure-supported quiescent galaxy or leave the rotation intact forming a rotating quiescent galaxy. In the second scenario, the disk-like kinematics and large disk circular velocities imply that compact quiescent galaxies would be observed as `fast rotators' at later times. Evidence of rotating quiescent galaxies at $z\gtrsim1$ from the literature includes observational results from deep imaging that suggest as many as $\gtrsim65$\% of compact quiescent galaxies are disk dominated at $z\sim2$ \citep{2008ApJ...677L...5V,2011ApJ...730...38V,2012ApJ...754L..24C} and spectroscopic results of two strongly lensed $z>2.1$ quiescent galaxies which reveal stellar rotation curves of $V_\mathrm{max}\approx190$ \kms and $V_\mathrm{max}\approx500$ \kms \citep{2015ApJ...813L...7N,2017Natur.546..510T}.  The rotation detected in both compact SFGs and quiescent galaxies implies that integrated $\sigma_*$ measurements of quiescent galaxies likely have a rotational component (see also \citealt{2017ApJ...834...18B}, Mendel et al.~\textit{in prep}). 

The dynamics of local descendants of rotating high-redshift compact galaxies depends strongly on their accretion histories. High-redshift compact galaxies may evolve into kinematically distinct cores, compact galaxies or S0s (e.g. \citealt{2016ARA&A..54..597C}). It is possible that they become the fast-rotators observed in high fractions locally ($50-95$\% of galaxies with $10<\log{M_*[\Msun]}<11$; e.g. \citealt{2011MNRAS.414..888E,2017MNRAS.472.1272V,2018ApJ...852...36G}). However, the stellar masses could still increase by almost an order of magnitude from $z\sim2-3$ to today (e.g. \citealt{2013ApJ...777...18M}). If the compact SFGs follow this path, they may loose angular momentum due to merging and evolve into slow rotators at the highest masses (e.g. $\log{M_*[\Msun]}>11$; \citealt{2016MNRAS.456.1030W}).

Merging is a possible mechanism to quench existing compact SFGs (e.g. \citealt{2015MNRAS.449..361W,2015MNRAS.450.2327Z}) which could either destroy or retain the existing rotational support \citep{2003ApJ...597..893N,2005A&A...430..115H,2010ApJ...722.1666W}. There are two possible major mergers (mass ratios of 1:1, 1:2) in our sample identified by spectroscopic redshifts and projected separations ($<300$ \kms, $<14$ kpc). However, the number of possible companions around compact SFGs within 1-sigma redshift errors and 50 kpc (drawing from the 3D-HST catalog) is consistent within the errors with the full SFG sample of massive galaxies, $\log{(M_*[\mathrm{\Msun}])}>10$, in \kmostdns, suggesting that major mergers are not the sole mechanism responsible for quenching compact SFGs.

%The first, a 1:1 merger, comprises two compact SFGs, COS4\_11363 and COS4\_11337 within 5.5 kpc as shown in Figure~\ref{fig.intspec1}. Both galaxies show broad \halpha emission and elevated \NII/H$\alpha$ with redshift difference of $\sim100$ \kms. The pair is identified with X-ray luminosity indicative of an AGN. HST imaging reveals faint tidal features suggestive of a past or ongoing interaction. The second possible 1:2 merger is between compact SFG GS4\_44066, also an X-ray detected AGN, and the more massive compact quiescent galaxy 10 kpc away with velocity separation of 280 \kms~(with spectroscopic redshift from the VIRIAL survey). Finally COS4\_06963 is undergoing an 1:3 merger with a less massive galaxy 13.5 kpc away with velocity separation 190 \kms~also detected in the KMOS data. 

AGN provide another mechanism to quench galaxies particularly at these masses and redshifts (e.g. \citealt{2005Natur.433..604D,2006ApJS..163....1H,2006MNRAS.365...11C,2015MNRAS.452.1841S}) and have been proposed specifically for compact SFGs as a likely quenching mechanism \citep{2013ApJ...765..104B}. With access to additional metrics to measure nuclear activity (NII/\halpha ratios, and deep data to recover broad underlying emission components) we have found an even higher rate of possible AGN activity in compact SFGs. We estimate the AGN incidence is {$\sim1.4\times$} higher in compact SFGs than the overall population at a fixed stellar mass. While we cannot rule out that this is partially a consequence of selection due to the emission from an AGN being attributed to star-formation or an AGN outshining a large underlying disk, it is suggestive that AGN may play a central role in the evolutionary tracks of compact SFGs.  The high \NII/\halpha emission line ratio in the compact SFG stack is consistent with emission line ratios found in quiescent galaxies both locally \citep{2006ApJ...648..281Y} and at high redshift \citep{2015ApJ...813L...7N,2017ApJ...841L...6B,2017Natur.546..510T}.

%%%%%%%%%%%%%%%%%%%%%%%%%%%%%%%%%%%%%%%%%%%%%%%%

%%%%%%%%%%%%%%%%
\section{Conclusions}
\label{sec.conc}
%%%%%%%%%%%%%%%%

We present the kinematic properties of {35} compact centrally-dense star-forming galaxies at $z=0.9-3.7$ in the \kmostd survey. For the first time with Integral Field Spectroscopy (IFS) we spatially resolve 23 compact SFGs. The IFS data map the emission line kinematics and morphology in two spatial dimensions enabling a determination of the kinematic position angle in rotating galaxies. The data reveal that the majority of resolved compact SFGs, {21/23}, are rotationally-dominated systems with rotational velocities and disk dispersions comparable to the full \kmostd dataset of SFGs at similar masses.  With the kinematic position angle know we can measure velocity gradients ranging from $95-500$ \kms. The integrated ionized gas linewidths of compact SFGs ($75-400$ \kms) and extended SFGs can be reproduced by a combination of their observed rotation, disk velocity dispersion, and wind strength.  This line decomposition demonstrates the important interplay of different kinematic components of these systems when inferring rotational velocities from unresolved data. %high AGN fraction

The \halpha profiles of compact SFGs are well fit with an exponential disk model with sizes of $\sim2$ kpc, slightly larger or comparable to the broad-band continuum sizes. This result and the detection of rotation are in line with earlier results presented by \cite{2015ApJ...813...23V}. Stacked spectra of compact SFGs show higher \NII/\halpha and comparable \SII/\halpha to a stellar mass and redshift matched sample. 

We derive large dynamical masses leaving little room for large molecular gas reservoirs $-$ a result supported by recent ALMA observations of two of the compact SFGs in our sample. The high M$_*$/M$_\mathrm{dyn}$ ratios together with the structural parameters (high central densities and cuspy profiles) suggest that, assuming no further gas replenishment, these galaxies will have their large SFRs quenched on a short time scale. Depending on the quenching mechanism it is possible that the resultant quiescent galaxies would retain the rotational support observed in the compact SFG phase.  There is a growing amount of evidence in the literature to support the scenario of rotationally supported quiescent galaxies at $z\sim0.5-3$. Integrated ionized gas linewidths of compact SFGs are comparable to the stellar velocity dispersions of compact quiescent galaxies at similar redshifts. However, direct measurement of stellar and gas velocity dispersions in the \textit{same objects} (e.g \citealt{2015arXiv150307164B}) for a large sample is required to see if a link between these two measurements can be made on average. Future work utilising synergies between the \kmostd survey and the complementary VIRIAL survey \citep{2015ApJ...804L...4M} of $UVJ$ passive galaxies in the same fields will investigate the relationship between stellar and ionized gas linewidths for a sample of galaxies observed in both surveys. 

%%%%%%%%%%%%%%%%%%%%%%%%%%%%%%%
\acknowledgments
%%%%%%%%%%%%%%%%%%%%%%%%%%%%%%%
We wish to thank the ESO staff, and in particular the staff at Paranal Observatory, for their helpful and enthusiastic support during the many observing runs over which the KMOS GTO were carried out. We thank the entire KMOS instrument and Commissioning team for their hard work. We also thank the software development team of \spark for all their work with us to get the most out of the data. DJW and MF acknowledge the support of the Deutsche Forschungsgemeinschaft via Project ID 3871/1-1 and  3871/1-2. EW acknowledges the support of ASTRO 3D funding for the writing retreat used to bring this paper to completion. Parts of this research were conducted by the Australian Research Council Centre of Excellence for All Sky Astrophysics in 3 Dimensions (ASTRO 3D), through project number CE170100013.

\bibliographystyle{apj}

%%%%%%%%%%%%%%%%%%%%%%%%%%%%%%%

%\clearpage

%%%%%%%%%%%%%%%%%%%%%%%%%%%%%%%%%%%%%%%%%%%%%%%%
\clearpage

\label{lastpage}

\end{document}